\documentclass[lettersize,journal]{IEEEtran}
\usepackage{amsmath,amsfonts}
\usepackage{amssymb} 
\DeclareMathAlphabet{\mathcal}{OMS}{cmsy}{m}{n}
\usepackage{txfonts}
\usepackage{graphicx}
\usepackage{cite}
\usepackage{booktabs,makecell}
\usepackage{algorithmic}
\usepackage{algorithm}
\usepackage{textcomp}
\usepackage{stfloats,color}

\newcommand{\larrow}{\overset{\scriptscriptstyle\leftarrow}}
\newcommand{\rarrow}{\overset{\scriptscriptstyle\rightarrow}}

\begin{document}

\title{Grant Free MIMO-NOMA with Differential Modulation for Machine Type Communications}
\author{Yuanyuan Zhang, Zhengdao Yuan, Qinghua Guo,~\IEEEmembership{Senior Member,~IEEE,} 
	Zhongyong Wang,  Jiangtao Xi, ~\IEEEmembership{Senior Member,~IEEE,} 
	Yanguang Yu and  Yonghui Li, ~\IEEEmembership{Fellow,~IEEE}
	\thanks{This work was supported in part by the National Natural Science Foundation of China under Grant 62101506, the Science and Technology Research Projects of Henan Province under Grant 222102210255 and 242102240127, and the Australian Research Council (ARC) Discovery Project under Grant DP210103410. (\textit{Yuanyuan	Zhang and Zhengdao Yuan contributed equally to this work. Corresponding authors: Qinghua Guo and Zhongyong Wang})  }
	\thanks{Y. Zhang  is with the School of Electronics and Information, Zhengzhou University of Light Industry, Zhengzhou 450002, Henan, China. She was with the School of Electrical, Computer and Telecommunications Engineering, University of Wollongong, Wollongong, NSW 2522, Australia (e-mail: ieyyzhang@zzuli.edu.cn). }
	\thanks{Z. Yuan is  with the Artificial Intelligence Technology Engineering Research Center, Open University of Henan, and School of Information Engineering, Zhengzhou University, Zhengzhou 450002, China (e-mail:yuan\_zhengdao@foximail.com).}
	\thanks{Q. Guo, J. Xi and Y. Yu are with the School of Electrical, Computer and Telecommunications Engineering, University of Wollongong, Wollongong, NSW 2522, Australia (e-mail: qguo@uow.edu.au; jiangtao@uow.edu.au; yanguang@uow.edu.au ).}
	\thanks{ Z. Wang  is with the School of Electrical and Information Engineering, Zhengzhou University, Zhengzhou 450001, China (e-mail: zywangzzu@gmail.com). }
	\thanks{ Y. Li is with the School of Electrical and Information Engineering, University of Sydney, Sydney, NSW 2006, Australia (e-mail: yonghui.li@sydney.edu.au).}
%	\thanks{Copyright (c) 20xx IEEE. Personal use of this material is permitted. However, permission to use this material for any other purposes must be obtained from the IEEE by sending a request to pubs-permissions@ieee.org. }
}

%\markboth{IEEE INTERNET OF THINGS JOURNAL,~Vol.~XX, No.~XX, Dec.~2021}%

\maketitle

\begin{abstract}
This paper considers a challenging scenario of machine type communications, where we assume internet of things (IoT) devices send short packets sporadically to an access point (AP) and the devices are not synchronized in the packet level. High transmission efficiency and low latency are concerned. Motivated by the great potential of multiple-input multiple-output non-orthogonal multiple access (MIMO-NOMA) in massive access, we design a grant-free MIMO-NOMA scheme, and in particular differential modulation is used so that expensive channel estimation at the receiver (AP) can be bypassed. The receiver at AP needs to carry out active device detection and multi-device data detection. The active user detection is formulated as the estimation of the common support of sparse signals, and a message passing based sparse Bayesian learning (SBL) algorithm is designed to solve the problem. Due to the use of differential modulation, we investigate the problem of non-coherent multi-device data detection, and develop a message passing based Bayesian data detector, where the constraint of differential modulation is exploited to drastically improve the detection performance, compared to the conventional non-coherent detection scheme. Simulation results demonstrate the effectiveness of the proposed active device detector and non-coherent multi-device data detector. 
\end{abstract}

\begin{IEEEkeywords}
 Grant-free, non-orthogonal multiple access (NOMA), internet of things (IoT), multiple-input multiple-output (MIMO), message passing, machine type communications. 
\end{IEEEkeywords}

\section{Introduction}

As an enabler for the Internet of Things (IoT), massive machine type communications have attracted much attention, which is characterized by massive connections and sporadic short-burst transmissions \cite{5Gscenarios2016,IOT2016}. Various IoT communication transmission scheduling mechanism have been studied in \cite{Wei1,Wei2,Wei3,Wei4}.
Due to the limited spectral resource, it is difficult to use the orthogonal multiple access (OMA) scheme to provide connectivity for a massive number of IoT devices. 
Nonorthogonal multiple access (NOMA), where a resource block is used to serve multiple users \cite{daiNOMA2015, NOMAIOT2020, NOMAIOT2021}, is considered to meet the demands on low latency and massive connectivity. The existing NOMA schemes can be mainly divided into two categories, i.e., power domain multiplexing such as basic power-domain NOMA \cite{PDNOMAupANDdown,PDnoma2020,PDNOMA2021,PDNOMA2021R2,PDNOMA2022R2}, and code domain multiplexing which includes low-density spreading (LDS) multiple access \cite{LDSOFDMmpa1}, sparse code multiple access (SCMA) \cite{SCMA2017,SCMA2021Liu}, etc.	
Moreover, in grant-based network, the handshaking between the receiver and the transmitter must be processed before the data transmission, i.e., data transmission requires a “Grant” \cite{GB2005}. Due to the handshaking procedure, the conventional grant-based transmission schemes may incur uncertain latency and excessive signaling overhead, which makes the communications inefficient, especially in the case that IoT devices have only a small amount of payload data, e.g., a few bits for each transmission \cite{MTC2014, mMTC2016}. Therefore, grant-free transmission without handshaking procedure, where users can randomly transmit data at any time instant, is highly desirable, which can save a huge amount of signaling overhead and reduce the communication latency to ensure the real-time communication of IoT, and that is very attractive for dynamic scenarios, e.g., in internet of vehicles (IoV). 
The grant-free non-orthogonal multiple access (NOMA)\cite{NOMAa,UADandMUD2019,iot2020,NOMAnew,Liu2022WCL}, which uses a resource block to serve multiple users and does not need the handshaking procedure, is a very promising technique for machine type communications.

\subsection{Recent Works and Motivation} 

Recently, to further enhance the spectrum efficiency, multiuser access  and/or communication reliability, NOMA has been combined with multiple-input and multiple-output (MIMO), leading to MIMO-NOMA\cite{MIMONOMAsum,PDnomaMIMOcapacity2,CDnomaMIMO}. The performance of MIMO-NOMA schemes  have been studied in \cite{PDnomaMIMOcapacity3,PDnomaMIMO3,mioNOMiot2022}.  
With the assumption that the channel state information (CSI) is available to the receiver, various signal processing techniques such as those based on Karush-Kuhn-Tucker (KKT) conditions\cite{ PDnomaMIMO4}, Gaussian message passing (GMP)\cite{ codeMIMONOMA3}, and improved Bayesian Learning \cite{ IBLmimoNOMA2023} have been studied. For practical applications, the CSI has to be acquired via training signals for coherent detection \cite{ PDnomaMIMOesmtCSI1,PDnomaMIMOesmtCSI2}, and  a MIMO-NOMA receiver based on deep learning is proposed in \cite{DLmimoNOMA2023}, where a deep neural network is implemented for the joint estimation of channel parameters and the detection of signals. 
However, due to the large number of variables to be estimated, the acquisition of accurate instantaneous CSI for (massive) MIMO leads to significant training overhead, and channel estimation is a challenging task, especially in the scenarios that wireless channels are fast time-varying. The implementation of grant-free MIMO-NOMA systems is even more challenging, but it has great intentional for massive machine type communications.

This paper deals with the design of grant-free MIMO-NOMA for machine type communications, where each device is equipped with a single antenna and the AP is equipped with a (massive) number of antennas. We assume that the data payload of the devices is small (e.g., a few bits), and the devices send information to an access point (AP) randomly. In addition, the data packet length of the devices is not fixed, so the devices are not synchronized in the packet level, i.e., any device is allowed to freely access the channel whenever it has information to transmit, and it becomes inactive when its packet is delivered. Due to the high flexibility of the transmissions and the consideration of high transmission efficiency, the realization of the system is very challenging, especially the implementation of the receiver at the AP, as it needs to identify the active devices and perform multi-device data detection. The conventional grant-free NOMA strategy is to first perform active user identification and channel estimation, followed by coherent data detection. However, due to a large number of antennas at the AP and the data packet of the devices is very short, the conventional strategy will render the transmissions highly inefficient due to the requirement of long training signals for active user identification and channel estimation. Even the large training overhead is not an issue, it is unknown how to use the conventional strategy to handle the asynchronous transmissions of the IoT devices, where the devices have different packet lengths and are allowed to access and leave the channel freely.

 \subsection{Contributions} 
The main contributions of this work are summarized below:

 (1) We propose to solve the challenge by adopting differential modulation at the devices and designing non-coherent multi-device data detection at the AP. Such design can bypass channel estimation. The implementation of the transmitters is very simple, where each device is assigned a spreading sequence and transmits information with differential modulation. The receiver needs to first identify active users without the assistance of any training signals, and then perform multi-device data detection for the active users. 

(2) We assume that the channels remain constant over two consecutive symbol intervals. The active user detection and multi-device data detection are performed symbol by symbol, so that the devices are allowed to access and leave the channel freely without packet level synchronization and channel estimation, which are the key to achieve high flexibility and high efficiency. To this end, the design of efficient active user detector and multi-device data detector is crucial. 

 (3) We use the factor graph and message passing techniques to address the problem. With the sparsity of active users, the active device detection is formulated as common support estimation of sparse signals, and we develop a message passing based sparse Bayesian learning (SBL) algorithm to solve it, where belief propagation (BP)  \cite{FactorG2001} and  mean field (MF) \cite{MF2002,VMP2005} are combined to implement the SBL. After active users are identified, non-coherent data detection needs to be performed. For this, we investigate the problem of non-coherent multi-device detection problem and design a BP-MF message passing based data detector, which treats the differential modulation as a constraint and encodes it into a probabilistic form as a local function in the factor graph representation. We show that exploiting the constraint can improve the detection performance remarkably, compared to the conventional detection scheme. Simulation results are provided to demonstrate the effectiveness of the proposed scheme and superior performance of the developed algorithms. 

The rest of this paper is organized as follows. In Section II, the grant free MIMO-NOMA system is designed and the problems of active device detection and non-coherent multi-device data detection are formulated. Message passing based non-coherent receiver is developed in Section III. Numerical simulation results are provided in Section IV, followed by conclusions in Section V.

\textit{Notation}- Lowercase and uppercase letters denote scalars. Boldface lowercase and uppercase letters denote column vectors and matrices, respectively. The superscriptions $(\cdot)^T $ and $(\cdot)^H $ denote the transpose and conjugate transpose operations, respectively, and $\varpropto $ denotes equality of functions up to a scale factor. The function $  \mathcal{CN}(x;\hat{x},\sigma^2_x) $  stands for  a complex Gaussian distribution with mean $\hat{x} $ and variance $\sigma^2_x $. As the convention, {\small $ \left< f\left( x,y,z\right) \right>_{f\left(y\right)f\left(z\right)}\text{=}\iint \!\! {f( x,y,z)f(y)f(z)dydz} $}  is used to denote the marginalization operator. The expectation operator with respect to a scaled probability density function (PDF) $ g(x) $ is expressed by {\small $\left<x\right>_{g\left(x\right)}\text{=}\int {xg(x)dx}/\int {g({{x}'})d{{x}'}} $}, and {\small $ \text{Var}\left[ x \right]_{g(x)} \text{=} \left< |x|^2 \right>_{g(x)} - |\left< x \right>_{g(x)}|^2 $} stands for the variance of $ x $.

\section{Grant Free MIMO-NOMA System Model and Problem Formulation}

\subsection{Grant Free MIMO-NOMA with Differential Modulation}

The grant free MIMO-NOMA system is illusrated in Fig. 1.   The key notations are listed in Table \ref{list1}.  
We assume an IoT communication system, where an access point (AP) equipped with $ N $ antennas is used to serve $ U $ IoT devices, each equipped with a single antenna. In particular, we consider uplink transmission, i.e., the IoT devices send information to the AP. As the transmissions of the IoT devices are sporadic, i.e., not all devices are active at the same time, we use  $ K $  to denote the number of active devices served by the AP.

\begin{figure}
	\centering
	\includegraphics[width=3in]{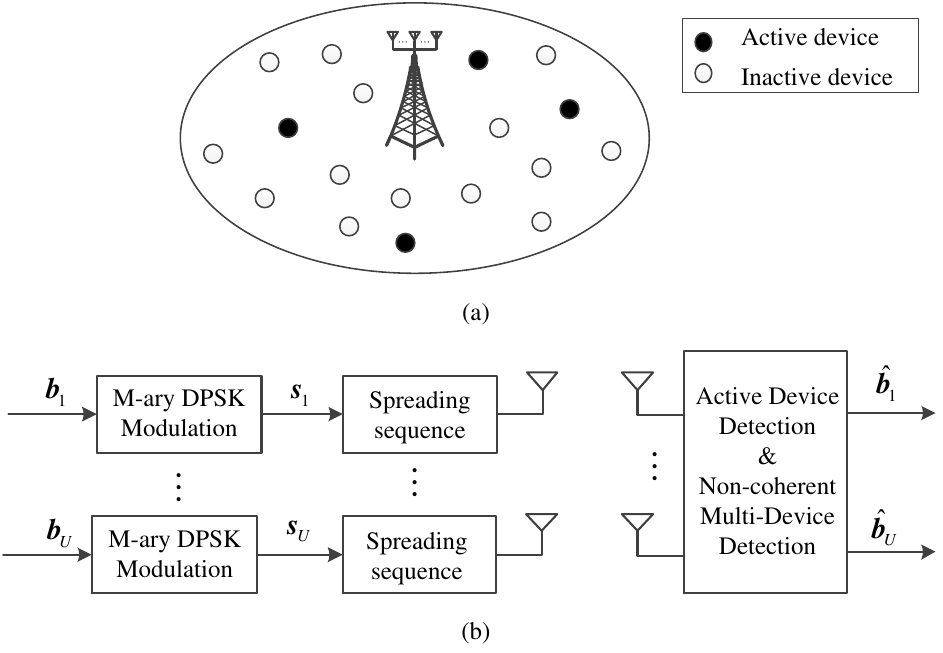}
	\caption{Illustration and system diagram of grant-free MIMO-NOMA.}
	\label{fig:BD}
\end{figure}

\begin{table}[t]
	\renewcommand\arraystretch{1.3} %设置为2倍行高
	\centering 
	\caption{ List of main symbols }\label{list1}
	 \begin{tabular}{|c|c|}	
			\hline
			\textbf{Notation} & \textbf{Description} \\ \hline 
			$ N $ & Number of antennas at AP \\ \hline
			$ n $ & Antenna index \\ \hline
			$ U $ & Total number of devices \\ \hline
			$ u $ & Device index \\ \hline
			$ K $ & Number of active devices \\ \hline
			$ k $ & Active device index \\ \hline
			$ T $ & Length of the symbol sequence  \\ \hline
			$ t $ &  Symbol index \\ \hline
			$ L $ & Length of spreading sequences (i.e. Number of subcarriers)  \\ \hline
			$ l $ &  Subcarrier index \\ \hline
			$ \mathcal{X}_d $ & DPSK demodulation symbol set  \\ \hline
	\end{tabular}   
\end{table}

Due to the small payload of the IoT devices, the requirement of low latency and (larger) number of antennas used at the AP, the use of training symbols for channel estimation and active device detection is not preferable. In this work, we design a non-coherent scheme and differential modulation is employed by the IoT devices, so that the use of training symbols can be avoided.
As show in Fig. \ref{fig:BD} (b), the input bit stream $ \textbf{b}_u $ of device $ u $ is mapped to a symbol sequence $  \textbf{s}_u = [s_{u,0},s_{u,1},\cdots,s_{u,T-1}]  $ with differential modulation, e.g.,  M-ary  differential phase shift keying (DPSK), where $ T $ is the length of the symbol sequence and $ s_{u,0} $ is a known reference symbol that is drawn from the M-ary DPSK constellation set $ \mathsf{\mathcal{X}} $. Then, each symbol in $ \textbf{s}_u $ is spread onto $ L $ chips using a unique spreading sequence $ \textbf{p}_u \in \mathbb{C}^{L \times 1}$ for device $ u $, which is a column of the spreading matrix $ \textbf{P}=[ \textbf{p}_1,\cdots,\textbf{p}_{_U}] $. To support a large number of users with spreading sequences of limited length, the spreading sequences are not orthogonal. In this paper, the Zadoff-Chu (ZC) sequences with different roots and shifts 
are employed as the spreading sequences.  
As not all users are active, we define $ A_u $ as the activity indicator for device $ u $, which takes 1 if the device is active, and 0 otherwise. The $ t $-th equivalent transmitted symbol of device $ u $ is denoted as 
\begin{equation}
\bar{s}_u ^{(t)}= A_u s_{u,t}.
\end{equation}
Then, the received signal at the $ n $-th antenna can be expressed as
\begin{equation}\label{yt}
\textbf{y}_n^{(t)}  = \textbf{P} \textbf{H}_n \bar{\textbf{s}}^{(t)}+\textbf{w}_n^{(t)},
\end{equation}
where $ \textbf{y}_n^{(t)} \in \mathbb{C}^{L \times 1} $,
$ \textbf{H}_n = diag ( \textbf{h}_n) $ is a diagonal matrix, whose diagonal elements are given as $ \textbf{h}_n = [h_{1,n},h_{2,n},\cdots ,h_{_U,n}] $, $h_{u,n} $ is the channel coefficient between device $ u $ and the $ n $-th antenna, which is complex Gaussian distributed with zero mean,
$ \bar{\textbf{s}}^{(t)} = [\bar{s}_1 ^{(t)} ,\bar{s}_2 ^{(t)} , \cdots, \bar{s}_U ^{(t)} ]^T$ is the $ t $-th transmitted equivalent symbol vector of length $U$, whose element $ \bar{s}_u ^{(t)} $  takes a value from the complex-constellation set  $ \bar{\mathsf{\mathcal{X}}} \triangleq\{ \mathsf{\mathcal{X}} \cup 0 \}$, and the noise vector on the $ n $-th antenna  $ \textbf{w}_{n}^{(t)} \sim   \mathcal{CN}(\textbf{w}_n^{(t)};0,\sigma^2_w\textbf{I}_{L})$. 

The received signals by $ N $ antennas can be expressed in a matrix form as
\begin{equation} \label{eq:Yt1}
\textbf{Y}^{(t)}  = \textbf{P} \left[\textbf{H}_1 \bar{\textbf{s}}^{(t)}, \textbf{H}_2 \bar{\textbf{s}}^{(t)}, ..., \textbf{H}_N \bar{\textbf{s}}^{(t)}\right] + \textbf{W}^{(t)},
\end{equation}
where $ \textbf{Y}^{(t)}= [\textbf{y}_1^{(t)}, \textbf{y}_2^{(t)} , \cdots , \textbf{y}_N^{(t)} ] $ is a matrix of size $ L \times N $, 
and $ \textbf{W} \in \mathbb{C}^{L\times N} $ is a matrix of white Gaussian noise whose power is unknown.

As mentioned before, the IoT devices are allowed to access and leave the channel freely as they may need to transmit their data to the AP randomly with varying packet length. By taking advantage of differential modulation, our aim is to obtain the transmitted information of active devices, while avoiding the estimation of the channel coefficients. To achieve this, both active device detection and data detection need to be carried out based on two consecutive received matrices $ \textbf{Y}^{(t)} $ and $ \textbf{Y}^{(t-1)}$, corresponding to two consecutive symbol intervals.

\subsection{Problem Formulation}
Define $ \textbf{x}_n^{(t)} = \textbf{H}_n \bar{\textbf{s}}^{(t)}$, where $ \textbf{x}_n^{(t)}=[x_{1,n}^{(t)},x_{2,n}^{(t)},\cdots,x_{_U,n}^{(t)}]^T $ is a vector of length $U$ and $ x_{u,n}^{(t)} = A_u h_{u,n} s_{u,t}$. From \eqref{eq:Yt1}, we have  
\begin{equation} \label{eq:Yt}
\textbf{Y}^{(t)}  = \textbf{P} \textbf{X}^{(t)} + \textbf{W}^{(t)},
\end{equation}
where $ \textbf{X}^{(t)}= [\textbf{x}_1^{(t)}, \textbf{x}_2^{(t)} , \cdots , \textbf{x}_{N}^{(t)} ] $ is an $ U \times N $ matrix. As joint active device detection and data detection (multi-device detection) will lead to high complexity, we adopt a two-stage strategy, i.e., performing active device identification followed by data detection.

\subsubsection{Active Device Detection}  
As only $ K $ out of $ U $ users are active, according to the definition of $ \textbf{X}^{(t)} $, we can see that each column of the matrix is a sparse vector, and all the columns share a common support. We need to find the non-zero rows of $ \textbf{X}^{(t)} $, then identify the spreading sequences of active users according to the indices of the non-zero rows (thereby active devices are detected). We define the common support of $ \textbf{X}^{(t)} $  as a set $ \mathsf{\mathcal{U}}_a $, whose elements are the indices of the non-zero rows of $ \textbf{X}^{(t)} $. Note that the cardinality of $ \mathsf{\mathcal{U}}_a $ is the number of active users, i.e., $ K = |\mathsf{\mathcal{U}}_a| $. Hence active device detection can be formulated as a problem of recovering sparse vectors with a common support, and we will design a sparse Bayesian learning algorithm to solve it. 

Based on the states of the devices over two consecutive symbol intervals, we can decide their activities. If a device is active at time $t-1$ while inactive at time $t$, the device finishes transmission. If a device is inactive at time $t-1$ while active at time $t$, the device just starts data transmission. If a device is active at both time $t-1$ and time $t$, the device is transmitting data, and we need to perform data detection for the device.

\subsubsection{Non-Coherent Multi-Device Data Detection}  
To facilitate multi-device data detection, we pick out the non-zero rows (whose indices belong to the support set $ \mathsf{\mathcal{U}}_a $) of $ \textbf{X}^{(t)} $ and $ \textbf{X}^{(t-1)} $, and the corresponding columns of $\textbf{P}$. Then (\ref{eq:Yt}) can be reduced to
\begin{equation} \label{eq:yt}
\textbf{Y}^{(t)}  = \bar{\textbf{P}} \bar{\textbf{X}}^{(t)} + \textbf{W}^{(t)},
\end{equation}
where $ \bar{\textbf{P}} $ is an  $ L\times K $ matrix, whose columns are drawn from  $ \textbf{P} $ based on the support set $ \mathsf{\mathcal{U}}_a $, $  \bar{\textbf{X}}^{(t)} $ is a $ K \times N $ matrix, whose rows are drawn from  $ \textbf{X}^{(t)} $  based on the support set $ \mathsf{\mathcal{U}}_a $.  

Assuming that DPSK is used, 
we define $ \psi_{k}^{(t)}= {s_{k,t}}/{s_{k,t-1}}$ to denote the $t$-th symbol of device $ k $. Then, we have 
\begin{align}
\psi_{k}^{(t)} \notag & =\frac{s_{k,t}}{s_{k,t-1}}\\
& = \frac{x_{k,1}^{(t)}}{x_{k,1}^{(t-1)}} = \frac{ h_{k,1} s_{k,t}}{ h_{k,1} s_{k,t-1}} \\
\notag &  = \cdots \\
\notag &= \frac{x_{k,N}^{(t)}}{x_{k,N}^{(t-1)}} = \frac{h_{k,N} s_{k,t}}{ h_{k,N} s_{k,t-1}}.
\end{align}
In a vector form, the $ t $-th symbol vector $ {\boldsymbol \psi}^{(t)} = [\psi_{1}^{(t)},\psi_{2}^{(t)},\cdots,\psi_{K}^{(t)}]^T$ over $ N $ antennas can be expressed as
\begin{align}\label{eq:diffSt} 
\notag & \underbrace{\left[\begin{array}{cccc}
	\psi_{1}^{(t)} & \psi_{1}^{(t)} & \cdots & \psi_{1}^{(t)}  \\
	\psi_{2}^{(t)} & \psi_{2}^{(t)} & \cdots & \psi_{2}^{(t)}  \\
	\vdots & \ddots & \ddots & \vdots \\
	\psi_{K}^{(t)} & \psi_{K}^{(t)} & \cdots & \psi_{K}^{(t)}  \\
	\end{array}\right]}_{K \times N}
=\bar{\textbf{X}}^{(t)} \oslash \bar{\textbf{X}}^{(t-1)} \\
& =\left[\begin{array}{cccc}
\frac{x_{1,1}^{(t)} }{ x_{1,1}^{(t-1)}}& \frac{ x_{1,2}^{(t)}}{x_{1,2}^{(t-1)} } & \cdots & \frac{ x_{1, N}^{(t)}}{x_{1, N}^{(t-1)} }  \\
\frac{x_{2,1}^{(t)}  }{ x_{2,1}^{(t-1)}}& \frac{ x_{2,2}^{(t)}}{ x_{2,2}^{(t-1)}} & \cdots & \frac{x_{2, N}^{(t)}}{x_{2, N}^{(t-1)}} \\
\vdots & \ddots & \ddots & \vdots \\
\frac{ x_{K, 1}^{(t)}}{ x_{K, 1}^{(t-1)}} & \frac{x_{K, 2}^{(t)} }{ x_{K, 2}^{(t-1)}} & \cdots & \frac{x_{K, N}^{(t)}}{x_{K, N}^{(t-1)}}
\end{array}\right],
\end{align}
where $ \oslash $ is the operation of element-wise division. 

The multi-device data detection is formulated as the estimation of $ {\boldsymbol \psi}^{(t)} $ based on $ \textbf{Y}^{(t)} $ and $ \textbf{Y}^{(t-1)}$. The conventional demodulation method carries out two steps:  first estimate $\bar{\textbf{X}}^{(t-1)}$ and $\bar{\textbf{X}}^{(t)}$ based on $ \textbf{Y}^{(t-1)}$ and $ \textbf{Y}^{(t)}$, respectively, then use \eqref{eq:diffSt} to estimate $\{\psi_{k}^{(t)}\}$, based on which hard decision for the information bits can be performed. In this paper, we propose a new method, where the decision on $\{\psi_{k}^{(t)}\}$ is obtained directly based on $\textbf{Y}^{(t-1)}$ and $ \textbf{Y}^{(t)}$ (instead of the intermediate results $\bar{\textbf{X}}^{(t-1)}$ and $\bar{\textbf{X}}^{(t)}$ as in the conventional two step approach), leading to remarkable performance gain as demonstrated later.

In this paper, both active device detection and multi-device data detection are solved with Bayesian methods. In particular, we use the factor graph techniques and design efficient message passing algorithms to solve the problems, which are elaborated in next section.

\section{ Message Passing Based Bayesian Approaches for Active Device Detection and Non-Coherent Multi-Device Data Detection} 
In this section, factor graph representations for the formulated problems are developed, based on which message passing-based Bayesian algorithms are designed for active device detection and multi-device data detection.

\subsection{Graph Representation for Block SBL-Based Active Device Detection}
To identify the active devices, we develop a block SBL method to estimate the common support of the row sparse matrix $ \textbf{X}^{(t)} $ in \eqref{eq:Yt}. Here, a two-layer hierarchical prior is employed, where the prior of the ($ n,u $)-th element in $ \textbf{X}^{(t)} $ is
\begin{align}\label{eq:SparsCon1}
p(x_{u, n}^{(t)})=\int p(x_{u, n}^{(t)}|\gamma_u) p(\gamma_u) d \gamma_u, 
\end{align}
where $ p(x_{u, n}^{(t)}|\gamma_u) = \mathcal{CN}(x_{u,n}^{(t)}; 0,\gamma_u^{-1} )  $, and  $ p(\gamma_u) = Ga(\gamma_u; \epsilon, \eta) $ with hyper-parameters $\epsilon$ and $\eta$. 
Due to the common support of the columns of $ \textbf{X}^{(t)} $, the elements of the $ u $-th row in $ \textbf{X}^{(t)} $ share a single hyper-parameter precision $ \gamma_u $, as shown in \eqref{eq:SparsCon1}. The elements of $ \textbf{X}^{(t)} $ are conditionally independent, and we have
\begin{align}
p(\textbf{X}^{(t)}| \boldsymbol{\gamma}) = \prod_{n=1}^{N}  \prod_{u=1}^{U}  p(x_{u,n}^{(t)}| \gamma_u), 
\end{align}
where $\boldsymbol{\gamma}=[\gamma_1, \gamma_2, ...,\gamma_U]^T$.
We assume that the noise precision at the $ n $-th antenna $ \lambda_n $ is unknown, and an improper prior $ p(\lambda_n) \propto 1/\lambda_n $ is assumed.

Then the joint distribution of $ \textbf{X}^{(t)}, \boldsymbol{\gamma}$ and $ \boldsymbol{\lambda} $ 
can be expressed as
\begin{align}\label{eq:pdfS5-1}
\notag & p(\textbf{X}^{(t)},\boldsymbol{\gamma},\boldsymbol{\lambda} | \textbf{Y}^{(t)}) \\
& \propto p(\textbf{Y}^{(t)}| \textbf{X}^{(t)},\boldsymbol{\lambda})  p(\textbf{X}^{(t)}| \boldsymbol{\gamma}) p(\boldsymbol{\lambda}) p(\boldsymbol{\gamma}) \\
\notag&= \prod_{u=1}^{U}  p(\gamma_u) \prod_{n=1}^{N}\Big(  \prod_{l=1}^{L}\left(  p(y_{l,n}^{(t)}| \textbf{x}_n^{(t)},\lambda_n) \right)   p(\lambda_n)  p(x_{u,n}^{(t)}| \gamma_u)  \Big) ,
\end{align}
where $ p(y_{l,n}^{(t)}| \textbf{x}_n^{(t)},\lambda_n) = \mathcal{CN}(y_{l,n}^{(t)}; \textbf{p}^T_l \textbf{x}_n^{(t)},\lambda_n^{-1})$ with $ \textbf{p}^T_l $ being the $ l $-th row of $ \textbf{P} $ and $ \textbf{x}_n^{(t)} $ being the $ n $-th column of $ \textbf{X}^{(t)} $ in \eqref{eq:Yt}, and  $\boldsymbol{\lambda}=[\lambda_1, ...,\lambda_N]^T$. 

\begin{figure}[!t]
	\centering
	\includegraphics[width=3.5in]{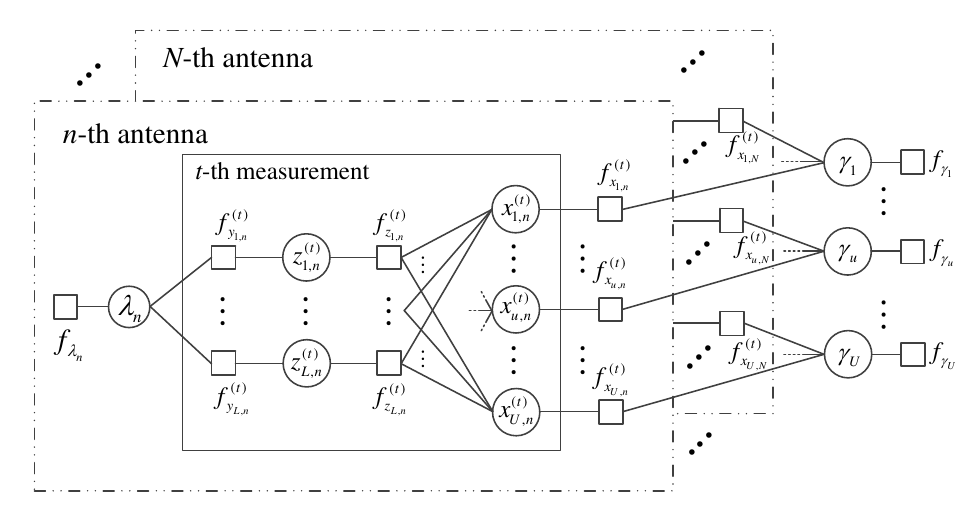}
	\caption{Factor graph representation of \eqref{eq:pdfS5-1}.}
	\label{FG5-1}
\end{figure}

To facilitate the factor graph representation of the factorization in \eqref{eq:pdfS5-1}, we introduce the notations in Table \ref{tabS5-1}, showing the correspondence between the factor labels and the underlying distributions they represent. 
In addition, to combine BP and MF message passing \cite{BPMF}, we introduce the auxiliary variable 
$ z_{l,n}^{(t)}= \textbf{p}^T_l \textbf{x}_n^{(t)}$ in Table \ref{tabS5-1},  inducing extra hard constrains $\delta (z_{l,n}^{(t)}- \textbf{p}^T_l \textbf{x}_n^{(t)}) $, which is denoted by $ f_{z_{_{l,n}}}^{(t)} (z_{l,n}^{(t)},\textbf{x}_n^{(t)}) $. Then $  p(y_{l,n}^{(t)}| \textbf{x}_n^{(t)},\lambda_n) $  becomes a function with  $ z_{l,n}^{(t)}  $ and $ \lambda_n $ as variables, i.e., $ \mathcal{CN}(y_{l,n}^{(t)};z_{l,n}^{(t)},\lambda_n^{-1})$. The factor graph representation of  (\ref{eq:pdfS5-1}) is shown in Fig. \ref{FG5-1}. The activity of device $ u $ is detected by comparing $ \hat\gamma_u $ with a threshold.

\begin{table}[t]
	\small
	\renewcommand\arraystretch{1.5} %设置为2倍行高
	\centering 
	\caption{The factors involved in the factorization in (\ref{eq:pdfS5-1}) }\label{tabS5-1}
	\begin{tabular}{l l l}
		\toprule[1pt] 
		\textbf{Factor} & \textbf{Distribution} & \textbf{Functional Form} \\
		\midrule[0.5pt]
		$ f_{\lambda_n}(\lambda_n) $ & $ p(\lambda_n) $ & $ 1/\lambda_n $ \\
		$ f_{y_{_{l,n}}}^{(t)}( z_{l,n}^{(t)},\lambda_n) $ & $ p(y_{l,n}^{(t)}| z_{l,n}^{(t)},\lambda_n) $ & $ \mathcal{CN}(y_{l,n}^{(t)};z_{l,n}^{(t)},\lambda_n^{-1}) $  \\
		$ f_{z_{_{l,n}}}^{(t)} (z_{l,n}^{(t)},\textbf{x}_n^{(t)})$  & $ p(z_{l,n}^{(t)}|\textbf{x}_n^{(t)}) $ &   $ \delta (z_{l,n}^{(t)}- \textbf{p}^T_l \textbf{x}_n^{(t)}) $\\
		$ f_{x_{u,n}}^{(t)}(x_{u,n}^{(t)},\gamma_u) $ & $ p(x_{u,n}^{(t)}| \gamma_u) $ & $ \mathcal{CN}(x_{u,n}^{(t)}; 0,\gamma_u^{-1} ) $\\
		%		$ f_{x_{u,n}}^{(t-1)}(x_{u,n}^{(t-1)},\gamma_u) $ & $ p(x_{u,n}^{(t-1)}| \gamma_u) $&  $ \mathcal{CN}(x_{u,n}^{(t-1)}; 0,\gamma_u^{-1} ) $\\
		$ f_{\gamma_u}(\gamma_u) $  & $ p (\gamma_u) $ & $ Ga(\gamma_u; \epsilon, \eta) $ \\
		\bottomrule[1pt]
	\end{tabular}
\end{table}

\subsection{Graph Representation for Bayesian Approach for Non-Coherent Multi-Device Data Detection}

With the results of active device detection, multi-device data detection is then carried out based on \eqref{eq:yt}. As differential modulation is used to avoid the training process, for the detection of the symbols of the active devices at time $t$, the received signal $\textbf{Y}^{(t-1)}$ at time $t-1$ is also required. In the design of our Bayesian approach, the differential modulation, i.e., the relationship between $ \textbf{X}^{(t)} $ and $ \textbf{X}^{(t-1)} $ in (\ref{eq:diffSt}) is regarded as a hard constraint. We transform it into a probabilistic form, and then the joint distribution of $ \bar{\textbf{X}}^{(t)} , \bar{\textbf{X}}^{(t-1)} $ and $  \boldsymbol{\psi}^{(t)} $ can be expressed as
\begin{align}\label{eq:DiffCon}
p( \bar{\textbf{X}}^{(t)} , \bar{\textbf{X}}^{(t-1)} , \boldsymbol{\psi} ^{(t)})
= \prod_{k=1}^{K} p(\psi_k^{(t)}) \prod_{n=1}^{N} p \Big(\bar{x}_{k,n}^{(t)},\bar{x}_{k,n}^{(t-1)} | \psi_k^{(t)} \Big),
\end{align}
where $ p(\bar{x}_{k,n}^{(t)},\bar{x}_{k,n}^{(t-1)} | \psi_k^{(t)}) $ represents a hard constraint, i.e.,  $ \delta (\bar{x}_{k,n}^{(t)} - \psi_k^{(t)} \bar{x}_{k,n}^{(t-1)} ) $, and $ p(\psi_k^{(t)}) $ is the prior of $ \psi_k^{(t)}$.  As $\psi_k^{(t)}$ is discrete valued, we have  $ p(\psi_k^{(t)}) = \sum_{q\in \mathsf{\mathcal{X}}_d}{\frac{1}{Q}\delta ( \psi_k^{(t)} -q )} $, and $ Q $ is the cardinality of the demodulation symbol set $ \mathsf{\mathcal{X}}_d $, i.e. $ Q= |\mathsf{\mathcal{X}}_d| $.
Given  $ \textbf{Y}^{(t)} $ and $ \textbf{Y}^{(t-1)} $, the joint a posteriori distribution of $ \bar{\textbf{X}}^{(t)},\bar{\textbf{X}}^{(t-1)}, \boldsymbol{\psi}^{(t)}$ and $ \boldsymbol{\lambda} $ 
can be expressed as
\begin{align}\label{eq:pdfS5-2}
\notag & p(\bar{\textbf{X}}^{(t)},\bar{\textbf{X}}^{(t-1)},\boldsymbol{\psi},\boldsymbol{\lambda} | \textbf{Y}^{(t)},\textbf{Y}^{(t-1)}) \\
\notag & \propto p(\textbf{Y}^{(t)}| \bar{\textbf{X}}^{(t)},\boldsymbol{\lambda}) p(\textbf{Y}^{(t-1)}| \bar{\textbf{X}}^{(t-1)},\boldsymbol{\lambda}) p(\bar{\textbf{X}}^{(t)},\bar{\textbf{X}}^{(t-1)},\boldsymbol{\psi})  p(\boldsymbol{\lambda}) \\
\notag &= \prod_{k=1}^{K} p(\psi_k^{(t)})  \prod_{n=1}^{N}\Bigg(  \prod_{l=1}^{L}\left(  p(y_{l,n}^{(t)}| \bar{\textbf{x}}_n^{(t)},\lambda_n) p(y_{l,n}^{(t-1)}| \bar{\textbf{x}}_n^{(t-1)},\lambda_n) \right)   \\
& \qquad \qquad \qquad \qquad \cdot  p(\lambda_n)   p\Big(\bar{x}_{k,n}^{(t)}, \bar{x}_{k,n}^{(t-1)}| \psi_k^{(t)}\Big)  \Bigg),  
\end{align}
where $ p(y_{l,n}^{(t)}| \bar{\textbf{x}}_n^{(t)},\lambda_n) = \mathcal{CN}(y_{l,n}^{(t)}; \bar{\textbf{p}}^T_l \bar{\textbf{x}}_n^{(t)},\lambda_n^{-1})$ with  $ \bar{\textbf{p}}^T_l $ being the $ l $-th row of $ \bar{\textbf{P}} $, and 
$ \bar{\textbf{x}}_n^{(t)} $ being the $ n $-th column of $ \bar{\textbf{X}}^{(t)} $  in (\ref{eq:yt}).

\begin{figure}[t]
	\centering
	\includegraphics[width=3.5in]{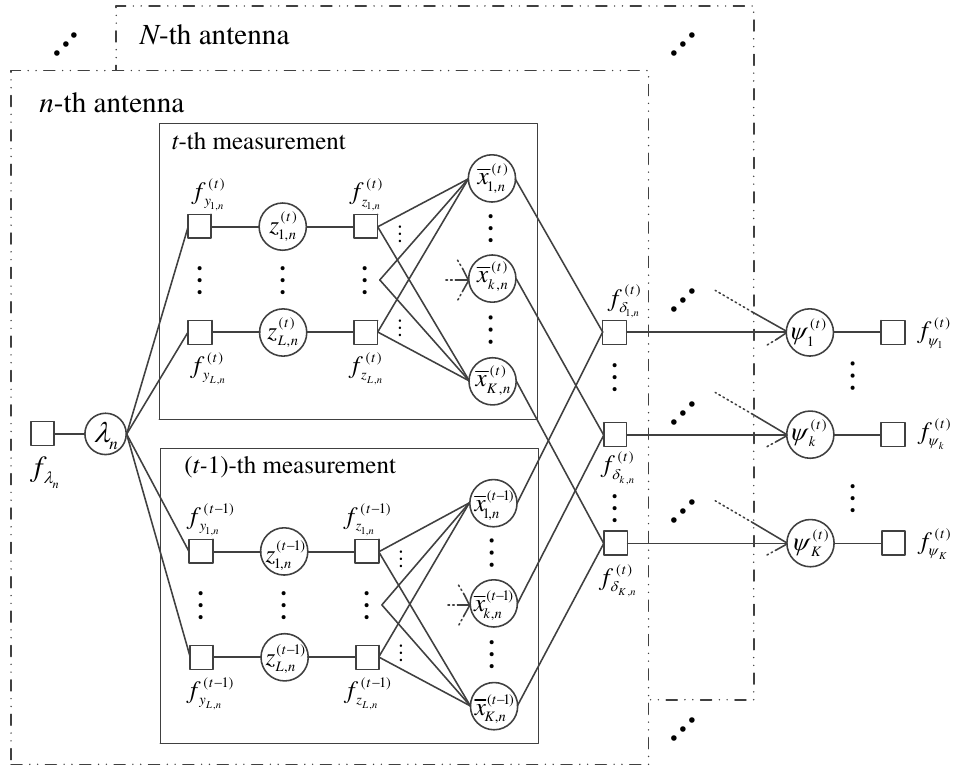}
	\caption{Factor graph representation of  (\ref{eq:pdfS5-2}). }
	\label{FG5-2}
\end{figure}

To facilitate the factor graph representation of the factorization in (\ref{eq:pdfS5-2}), we introduce the notations in Table \ref{tabS5-2}. 
In addition, we introduce the auxiliary variable 
$ z_{l,n}^{(t)}= \bar{\textbf{p}}^T_l \bar{\textbf{x}}_n^{(t)}$ in Table \ref{tabS5-2}, inducing extra hard constrains $ \delta( z_{l,n}^{(t)}- \bar{\textbf{p}}^T_l \bar{\textbf{x}}_n^{(t)}) $, which is denoted by $ f_{z_{_{l,n}}}^{(t)} ( z_{l,n}^{(t)},\bar{\textbf{x}}_n^{(t)}) $.
The stretched factor graph representation of (\ref{eq:pdfS5-2}) is shown in Fig.\ref{FG5-2}. 
It is noted that the size of the factor graph in Fig.\ref{FG5-2} is much smaller than that in Fig.\ref{FG5-1} because the inactive users have been removed, leading to significant advantage in terms of computational complexity, compared to the strategy of joint active device detection and data detection.

\begin{table}[t]
	\renewcommand\arraystretch{1.4} %设置为2倍行高
	\centering 
	\caption{The factors involved in the factorization in (\ref{eq:pdfS5-2}) }\label{tabS5-2}
	\resizebox{\linewidth}{!}{
		\begin{tabular}{l l l}
			\toprule[1pt] 
			\textbf{Factor} & \textbf{Distribution} & \textbf{Functional Form} \\
			\midrule[0.5pt]
			$ f_{\lambda_n}(\lambda_n) $ & $ p(\lambda_n) $ & $ 1/\lambda_n $ \\
			$ f_{y_{_{l,n}}}^{(t)}(z_{l,n}^{(t)},\lambda_n) $ & $ p(y_{l,n}^{(t)}| z_{l,n}^{(t)},\lambda_n) $ & $ \mathcal{CN}(y_{l,n}^{(t)}; z_{l,n}^{(t)},\lambda_n^{-1}) $  \\
			$ f_{ z_{_{l,n}}}^{(t)} (z_{l,n}^{(t)},\bar{\textbf{x}}_n^{(t)})$  & $ p( z_{l,n}^{(t)}|\bar{\textbf{x}}_n^{(t)}) $ &   $ \delta ( z_{l,n}^{(t)}- \bar{\textbf{p}}^T_l \bar{\textbf{x}}_n^{(t)}) $\\
			$ f_{\delta_{k,n}}^{(t)} (\bar{x}_{k,n}^{(t)},\bar{x}_{k,n}^{(t-1)},\psi_k^{(t)})$ & $ p(\bar{x}_{k,n}^{(t)},\bar{x}_{k,n}^{(t-1)}|\psi_k^{(t)}) $ & $ \delta (\bar{x}_{k,n}^{(t)} - \psi_k^{(t)} \bar{x}_{k,n}^{(t-1)} ) $ \\
			$ f_{\psi_k^{(t)}}(\psi_k^{(t)}) $  & $ p(\psi_k^{(t)}) $ & $ \sum\limits_{q\in \mathsf{\mathcal{X}_d}}{\frac{1}{Q}\delta ( \psi_k^{(t)} -q )} $  \\
			\bottomrule[1pt]
		\end{tabular}
	}
\end{table}

\subsection{Message Passing Algorithm for Active Device Detection  }

We detail the forward (from left to right) and backward (from right to left) message computations at each node of Fig. \ref{FG5-1} for active device detection, where some approximations are introduced to reduce the computational complexity. We use $ I_{A\to B}(x) $ to denote a message passed from a variable node (function node) $ A $ to a function node (variable node) $ B $, which is a function of $ x $.  The notations $ m $ and $ v $ are used to denote the mean and variance of a Gaussian message specified by their subscripts. The arrows over $ m $ and   $ v $ represent the directions of Gaussian massage passing. 
Note that, if a forward message computation requires backward  messages, we use the messages in previous iteration by default.

In  Fig. \ref{FG5-1}, combined BP and MF based message passing is used to realize the estimation of 
$ \{  \gamma_{u} , \forall u\} $, where the function nodes  $ \{ f_{y_{_{l,n}}}^{(t)}( z_{l,n}^{(t)},\lambda_n),\forall l,n\},$  $ \{f_{y_{_{l,n}}}^{(t-1)}( z_{l,n}^{(t-1)},\lambda_n),\forall l,n\},$  $\{ f_{x_{u,n}}^{(t)}(x_{u,n}^{(t)},\gamma_u),$  $\forall u,n\} $  and $ \{ f_{x_{u,n}}^{(t-1)}(x_{u,n}^{(t-1)},\gamma_u),\forall u,n\} $ are handled by the MF rule, and the function nodes  $ \{ f_{z_{_{l,n}}}^{(t)}, \forall l,n \} $ and $ \{ f_{z_{_{l,n}}}^{(t-1)}, \forall l,n \} $ are handled by the BP rules. We note that some proper message approximations are used to reduce the computational complexity.

\subsubsection{Forward message computations}

With the belief $ b\left(\lambda_{n}\right) $ whose computation is delayed to (\ref{eq:blfLmdS5}), the outgoing forward message at $ f_{y_{_{l,n}}}^{(t)} $ can be expressed as
\begin{align}\label{eq:fy2zS5}
\notag I_{f_{y_{_{l,n}}}^{(t)} \rightarrow z_{l, n}^{(t)}}\big(z_{l, n}^{(t)}\big) & \propto \exp \left\{\int \ln \left[f_{y_{_{l,n}}}^{(t)}\left(\lambda_{n}, z_{l, n}^{(t)}\right)\right] b\left(\lambda_{n}\right) d \lambda_{n}\right\} \\
& \propto \mathcal{CN} \left(z_{l, n}^{(t)} ; y_{l, n}^{(t)}, \hat{\lambda}_{n}^{-1}\right),
\end{align}
where $ {\hat{\lambda_n }} $ is calculated in (\ref{eq:LmdHat}). With {\small $ I_{z_{l, n}^{(t)} \rightarrow f_{z_{_{l,n}}}^{(t)} }\big(z_{l, n}^{(t)}\big) = I_{f_{y_{_{l,n}}}^{(t)} \rightarrow z_{l, n}^{(t)}}\big(z_{l, n}^{(t)}\big) $} and {\small  $ \{ I_{x_{u', n}^{(t)} \rightarrow f_{z_{_{l,n}}}^{(t)} }\big(x_{u', n}^{(t)}\big) , \forall u' \neq u\} $} computed in (\ref{eq:x2fxS5}), the outgoing forward message at $ f_{z_{_{l,n}}}^{(t)} $ can be expressed as
\begin{align}\label{eq:fz2xS5}
\notag I_{f_{z_{_{l,n}}}^{(t)} \rightarrow x_{u, n}^{(t)} }\big(x_{u, n}^{(t)}\big)
& = \int f_{z_{_{l,n}}}^{(t)} \big(z_{l,n}^{(t)},\textbf{x}_n^{(t)} \big)  I_{z_{l, n}^{(t)} \rightarrow f_{z_{_{l,n}}}^{(t)} }\big(z_{l, n}^{(t)}\big)\\
\notag & \qquad \quad \cdot \prod_{u' \neq u} I_{x_{u', n}^{(t)} \rightarrow f_{z_{_{l,n}}}^{(t)} }\big(x_{u', n}^{(t)}\big) d z_{l, n}^{(t)} d\textbf{x}_n^{(t)} \backslash x_{u, n}^{(t)}\\
&\propto \mathcal{CN}\left( x_{u, n}^{(t)}; \overset{\scriptscriptstyle\twoheadrightarrow}{m}_{x_{u, n,l}^{(t)}}, \overset{\scriptscriptstyle\twoheadrightarrow}{v}_{x_{u, n,l}^{(t)}} \right),
\end{align}
where
\begin{align}\label{eq:fz2xM}
\overset{\scriptscriptstyle\twoheadrightarrow}{m}_{x_{u, n,l}^{(t)}} =\frac{y_{l, n}^{(t)}-\larrow{m}_{z_{l, n}^{(t)}} + P_{_{l, u}} \overset{\scriptscriptstyle\twoheadleftarrow}{m}_{x_{u, n,l}^{(t)}} }{P_{_{l, u}}}, 
\end{align}
\begin{align}\label{eq:fz2xV}
\overset{\scriptscriptstyle\twoheadrightarrow}{v}_{x_{u, n,l}^{(t)}} = \frac{\hat{\lambda}_{n}^{-1} + \larrow{v}_{z_{l, n}^{(t)}}-\left|P_{_{l, u}}\right|^{2} \overset{\scriptscriptstyle\twoheadleftarrow}{v}_{x_{u, n,l}^{(t)}} }{\left|P_{_{l, u}}\right|^{2}},
\end{align}
and
\begin{align}
\larrow{m}_{z_{l, n}^{(t)}} = \sum_{u=1}^{U} P_{_{l, u}} \overset{\scriptscriptstyle\twoheadleftarrow}{m}_{x_{u, n,l}^{(t)}},
\end{align}
\begin{align}
\larrow{v}_{z_{l, n}^{(t)}} = \sum_{u=1}^{U} \left|P_{_{l, u}}\right|^{2} \overset{\scriptscriptstyle\twoheadleftarrow}{v}_{x_{u, n,l}^{(t)}}.
\end{align}
Thus, with {\small  $ \{ I_{f_{z_{_{l,n}}}^{(t)} \rightarrow x_{u, n}^{(t)} }\big(x_{u, n}^{(t)}\big) , \forall l \} $}, the outgoing forward message at $ x_{u, n}^{(t)} $ reads
\begin{align}
\notag I_{x_{u, n}^{(t)} \rightarrow f_{x_{u, n}}^{(t)}}\big(x_{u, n}^{(t)}\big)  & = \prod_{l=1}^L I_{f_{z_{_{l, n}}}^{(t)} \rightarrow x_{u, n}^{(t)}}\big(x_{u, n}^{(t)}\big) \\
& \propto \mathcal{CN} \left(x_{u, n}^{(t)} ; \rarrow{m}_{x_{u, n}^{(t)}}, \rarrow{v}_{x_{u,n}^{(t)}}\right),
\end{align}
where
\begin{align}\label{eq:x2fxV}
\rarrow{v}_{x_{u, n}^{(t)}}  =\left(\sum_{l=1}^L \frac{ 1}{\overset{\scriptscriptstyle\twoheadrightarrow}{v}_{x_{u, n,l}^{(t)}} }  \right)^{-1} \approx \left(\sum_{l=1}^L \frac{\left|P_{_{l, u}}\right|^{2}}{\hat{\lambda}_{n}^{-1}+\larrow{v}_{z_{l, n}^{(t)}}}\right)^{-1}, 
\end{align}
\begin{align}\label{eq:x2fxM}
\rarrow{m}_{x_{u, n}^{(t)}} \approx \rarrow{v}_{x_{u, n}^{(t)}} \sum_{l=1}^L \frac{P_{_{l, u}}^{H}\left(y_{l, n}^{(t)}-\larrow{m}_{z_{l, n}^{(t)}}\right)}{\hat{\lambda}_{n}^{-1}+\larrow{v}_{z_{l, n}^{(t)}}} + {m}_{x_{u,n}^{(t)}},
\end{align}
and the derivation of (\ref{eq:x2fxM}) is provided in Appendix A.
With {\small $ I_{x_{u, n}^{(t)} \rightarrow f_{x_{u, n}}^{(t)}}\big(x_{u, n}^{(t)}\big) $} and {\small $ I_{f_{x_{u, n}}^{(t)} \rightarrow x_{u, n}^{(t)}}\big( x_{u, n}^{(t)} \big) $} given in (\ref{eq:fx2x}), the belief of  $ x_{u, n}^{(t)} $ can be updated as
\begin{align}\label{eq:blfxS5}
\notag b \big(x_{u, n}^{(t)}\big) &= I_{x_{u, n}^{(t)} \rightarrow f_{x_{u, n}}^{(t)}}\big(x_{u, n}^{(t)}\big)   I_{f_{x_{u, n}}^{(t)} \rightarrow x_{u, n}^{(t)}}\big(x_{u, n}^{(t)}\big)\\
& \propto \mathcal{CN} \left( x_{u, n}^{(t)} ;  {m}_{x_{u, n}^{(t)}}, {v}_{x_{u,n}^{(t)}} \right),
\end{align}
where
\begin{align}\label{eq:mx} 
{m}_{x_{u,n}^{(t)}} = \frac{ \rarrow{m}_{x_{u, n}^{(t)}} }{ 1+ \hat{\gamma}_u \rarrow{v}_{x_{u, n}^{(t)}} },
\end{align}
\begin{align}\label{eq:vx} 
{v}_{x_{u, n}^{(t)}}= \left(  \frac{1}{\rarrow{v}_{x_{u, n}^{(t)}}}   + \hat{\gamma}_u  \right)^{-1}, 
\end{align}
and $ \hat{\gamma}_u $ is given by (\ref{eq:gammaHat}). Thus, the outgoing forward message at $ f_{x_{u, n}}^{(t)} $ can be computed as
\begin{align}
\notag I_{f_{x_{u, n}}^{(t)} \rightarrow \gamma_u}\big(\gamma_u\big) 
& \propto \exp \left\{\int \ln \left[f_{x_{u, n}}^{(t)}\left( x_{u, n}^{(t)} , \gamma_u \right)\right] b\left( x_{u, n}^{(t)} \right) d  x_{u, n}^{(t)} \right\} \\
& \propto  \gamma_u \exp  \left\{  - \gamma_u \left(   \left| {m}_{x_{u, n}^{(t)}} \right|^2    + {v}_{x_{u, n}^{(t)}} \right) \right\}.
\end{align}
With {\small $ \{ I_{f_{x_{u, n}}^{(t)} \rightarrow \gamma_u}\big(\gamma_u\big), \forall u,n\} $ } and  $ I_{f_{\gamma_u} \rightarrow \gamma_u}\big(\gamma_u\big) $ computed in (\ref{eq:fGtoG}), the belief of  $ \gamma_u $ can be updated as
\begin{align}
	\notag b \big(\gamma_u\big)  & = I_{f_{\gamma_u} \rightarrow \gamma_u}\big(\gamma_u\big) \prod_{n=1}^{N} I_{f_{x_{u, n}}^{(t)} \rightarrow \gamma_u}\big(\gamma_u\big)   \\
	& \propto  \gamma_u^{\hat{\epsilon} + N-1} \exp  \left\{ \!\! - \gamma_u\Big[ \eta + \sum_{n=1}^{N} \Big(   \big| {m}_{x_{u, n}^{(t)}} \big|^2    + {v}_{x_{u, n}^{(t)}}  \Big) \Big]  \right\}.
\end{align}
Thus, the estimation of $ \gamma_u $ can be updated by
\begin{align}\label{eq:gammaHat}
	\hat{\gamma}_u 
	={\left\langle \gamma_u  \right\rangle}_{b \big(\gamma_u\big)}
	=\frac{\hat{\epsilon}+N}{ \eta + \sum\limits_{n=1}^{N} \left(   \big| {m}_{x_{u, n}^{(t)}} \big|^2    + {v}_{x_{u, n}^{(t)}}  \right) }.
\end{align}
According to the results in \cite{UAMPSBL}, to enhance the performance of SBL, we can set $ \eta =0 $, and update $\epsilon $ automatically, i.e.,  
\begin{align}\label{eq:Ehat}
	\hat{\epsilon}& = \frac{1}{2}\sqrt{ \log \big(\frac{1}{U} \sum_{u} \hat{\gamma}_{u}\big)-\frac{1}{U} \sum_{u} \log \hat{\gamma}_{u}}.
\end{align}

\subsubsection{Backward message computations}
According to the MF rule, the outgoing backward message at $ f_{\gamma_u} $  can be calculated by
\begin{align}\label{eq:fGtoG}
I_{f_{\gamma_u} \rightarrow \gamma_u}\big(\gamma_u\big) 
& \propto Ga(\gamma_u; \hat{\epsilon}, \eta) .
\end{align}
With $ b \big(\gamma_u\big) $, the outgoing backward message at $ f_{x_{u, n}}^{(t)} $ can be computed as
\begin{align}\label{eq:fx2x}
\notag I_{f_{x_{u, n}}^{(t)} \rightarrow x_{u, n}^{(t)} }\big( x_{u, n}^{(t)} \big) 
& \propto \exp \left\{\int \ln \left[f_{x_{u, n}}^{(t)}\left( x_{u, n}^{(t)} , \gamma_u \right)\right] b\left( \gamma_u \right) d \gamma_u \right\} \\
& \propto   \mathcal{CN} \left( x_{u, n}^{(t)} ;  0, \hat{\gamma}_u^{-1} \right).
\end{align}
With the calculated {\small $ I_{f_{z_{_{l,n}}}^{(t)} \rightarrow x_{u, n}^{(t)} }\big( x_{u, n}^{(t)} \big) $} in (\ref{eq:fz2xS5}) and $ b \big(x_{u, n}^{(t)}\big) $  in (\ref{eq:blfxS5}), the outgoing backward message at  $ x_{u, n}^{(t)}  $  is given as
\begin{align}\label{eq:x2fxS5}
\notag I_{x_{u, n}^{(t)} \rightarrow f_{z_{_{l,n}}}^{(t)} }\big(x_{u, n}^{(t)}\big) &= \frac{b \big(x_{u, n}^{(t)}\big)}{  I_{f_{z_{_{l,n}}}^{(t)} \rightarrow x_{u, n}^{(t)}}\big(x_{u, n}^{(t)}\big) }  \\
& \propto \mathcal{CN} \left( x_{u, n}^{(t)} ; \overset{\scriptscriptstyle\twoheadleftarrow}{m}_{x_{u, n,l}^{(t)}}, \overset{\scriptscriptstyle\twoheadleftarrow}{v}_{x_{u, n,l}^{(t)}} \right),
\end{align}
where
\begin{align}\label{eq:x2fzV}
\overset{\scriptscriptstyle\twoheadleftarrow}{v}_{x_{u, n,l}^{(t)}}=\left( \frac{1}{ {v}_{x_{u, n}^{(t)}} } - \frac{1}{ \overset{\scriptscriptstyle\twoheadrightarrow}{v}_{x_{u, n,l}^{(t)}} } \right) ^{-1} \approx {v}_{x_{u, n}^{(t)}},
\end{align}
and 
\begin{align}\label{eq:x2fzM}
\notag \overset{\scriptscriptstyle\twoheadleftarrow}{m}_{x_{u, n,l}^{(t)}} & =  \overset{\scriptscriptstyle\twoheadleftarrow}{v}_{x_{u, n,l}^{(t)}}   \left( \frac{ {m}_{x_{u,n}^{(t)}} }{ {v}_{x_{u, n}^{(t)}} } - \frac{\overset{\scriptscriptstyle\twoheadrightarrow}{m}_{x_{u, n,l}^{(t)}}}{ \overset{\scriptscriptstyle\twoheadrightarrow}{v}_{x_{u, n,l}^{(t)}} } \right) \\
& \overset{(\ref{eq:x2fzV})}{\approx}  {m}_{x_{u, n}^{(t)}} -  {v}_{x_{u, n}^{(t)}} \frac{\overset{\scriptscriptstyle\twoheadrightarrow}{m}_{x_{u, n,l}^{(t)}}}{ \overset{\scriptscriptstyle\twoheadrightarrow}{v}_{x_{u, n,l}^{(t)}} }.
\end{align}
With {\small $ \{ I_{x_{u, n}^{(t)} \rightarrow f_{z_{_{l,n}}}^{(t)} }\big(x_{u, n}^{(t)}\big) ,\forall u \} $},  the outgoing backward message at $ f_{z_{_{l,n}}}^{(t)} $ can be expressed as
\begin{align}\label{eq:fz2zS5}
\notag I_{f_{z_{_{l,n}}}^{(t)} \rightarrow z_{l,n}^{(t)} }\big( z_{l,n}^{(t)}\big)
& = \int f_{z_{_{l,n}}}^{(t)} \big(z_{l,n}^{(t)},\textbf{x}_n^{(t)} \big)    \prod_{u = 1}^{U} I_{x_{u, n}^{(t)} \rightarrow f_{z_{_{l,n}}}^{(t)} }\big(x_{u, n}^{(t)}\big)  d\textbf{x}_n^{(t)} \\
&\propto \mathcal{CN}\left( z_{l,n}^{(t)}; \larrow{m}_{z_{l,n}^{(t)}}, \larrow{v}_{z_{l,n}^{(t)}} \right),
\end{align}
where
\begin{align}\label{eq:fz2zV} 
\larrow{v}_{z_{l,n}^{(t)}} = \sum_{u = 1}^{U} \left|P_{_{l, u}}\right|^{2} \overset{\scriptscriptstyle\twoheadleftarrow}{v}_{x_{u, n,l}^{(t)}}  \overset{(\ref{eq:x2fzV})}{\approx} \sum_{u = 1}^{U} \left|P_{_{l, u}}\right|^{2} {v}_{x_{u, n}^{(t)}} ,  
\end{align} 
\begin{align}\label{eq:fz2zM} 
\larrow{m}_{z_{l,n}^{(t)}} \approx \sum_{u = 1}^{U} P_{_{l, u}} {m}_{x_{u, n}^{(t)}}  - \frac{\larrow{v}_{z_{l,n}^{(t)}} \big( y_{l, n}^{(t)}- {^{i-1} \larrow{m}_{z_{l, n}^{(t)}}} \big) }{ \hat{\lambda}_{n}^{-1} + {^{i-1} \larrow{v}_{z_{l, n}^{(t)}} }},
\end{align} 
and the derivation of (\ref{eq:fz2zM})  is provided in Appendix A.
Note that, to update $ \larrow{m}_{z_{l,n}^{(t)}} $, we need $ \larrow{v}_{z_{l,n}^{(t)}} $ and $ \larrow{m}_{z_{l,n}^{(t)}} $ in the previous iteration, so the superscript $i-1$ is introduced to differentiate them.
With {\small $ I_{f_{z_{_{l,n}}}^{(t)} \rightarrow z_{l,n}^{(t)} }\big( z_{l,n}^{(t)}\big) $ } and {\small $ I_{f_{y_{_{l,n}}}^{(t)} \rightarrow z_{l, n}^{(t)}}\big(z_{l, n}^{(t)}\big) $} computed in (\ref{eq:fy2zS5}), the belief of $ z_{l, n}^{(t)} $ can be updated as
\begin{align}\label{eq:Blfz}
\notag b \big(  z_{l, n}^{(t)} \big) & = I_{f_{y_{_{l,n}}}^{(t)} \rightarrow z_{l, n}^{(t)}}\big(z_{l, n}^{(t)}\big)   I_{ f_{z_{_{l,n}}}^{(t)}  \rightarrow z_{l, n}^{(t)}}\big(z_{l, n}^{(t)}\big) \\
& \propto  \mathcal{CN}\left(  z_{l,n}^{(t)};  {m}_{z_{l,n}^{(t)}},  {v}_{z_{l,n}^{(t)}} \right) ,
\end{align}
where
\begin{align} 
\label{eq:BlfzV} {v}_{z_{l,n}^{(t)}} = \left( \hat{\lambda}_{n}  + \frac{1}{ \larrow{v}_{z_{l,n}^{(t)}} } \right) ^{-1} ,
\end{align}
\begin{align} 
\label{eq:BlfzM} {m}_{z_{l,n}^{(t)}} = {v}_{z_{l,n}^{(t)}} \left( y_{l, n}^{(t)} \hat{\lambda}_{n} + \frac{ \larrow{m}_{z_{l,n}^{(t)}} }{ \larrow{v}_{z_{l,n}^{(t)}} } \right).
\end{align}
Thus, the outgoing backward message at  $ f_{y_{_{l,n}}}^{(t)} $ is given by
\begin{align}\label{eq:fy2lmdS5}
\notag I_{f_{y_{_{l,n}}}^{(t)} \rightarrow \lambda_{n} }\big( \lambda_{n} \big) & \propto \exp \left\{\int \ln \left[f_{y_{_{l,n}}}^{(t)}\left(\lambda_{n}, z_{l, n}^{(t)} \right)\right]  b\left(  z_{l, n}^{(t)} \right) d  z_{l, n}^{(t)}\right\} \\
& \propto  \lambda_{n} \exp \left\{ - \lambda_{n} \left(  {{\Big|  {m}_{z_{l,n}^{(t)}}  -  y_{l, n}^{(t)} \Big|}^{2}} +  {v}_{z_{l,n}^{(t)}} \right)  \right\} .
\end{align}
With  $ \{  I_{f_{y_{_{l,n}}}^{(t)} \rightarrow \lambda_{n} }\big( \lambda_{n} \big)   ,\forall l \} $ and the   a priori $  f_{\lambda_n }\left( \lambda_n \right) $, the belief of $ \lambda_n $ can be updated by
\begin{align}\label{eq:blfLmdS5}
\notag b \big( \lambda_{n} \big)  & = f_{\lambda_n }\left( \lambda_n \right) \prod_{l = 1}^{L} I_{f_{y_{_{l,n}}}^{(t)} \rightarrow \lambda_{n} }\big( \lambda_{n} \big)   \\
& \propto \lambda_{n}^{L-1} \exp \Big\{ \!\! - \lambda_{n}   \sum_{l = 1}^{L} \Big( \big|  {m}_{z_{l,n}^{(t)}}  -  y_{l, n}^{(t)} \big|^{2} +  {v}_{z_{l,n}^{(t)}}  \Big)   \Big\}.
\end{align}
Thus, the estimation of noise precision is given as 
\begin{align}\label{eq:LmdHat}
\hat{\lambda}_n 
={\left\langle \lambda_{n}  \right\rangle}_{b_{\lambda_{n}} \big( \lambda_{n} \big)} 
=\frac{ L}{  \sum\limits_{l = 1}^{L} \Big(  \big|  {m}_{z_{l,n}^{(t)}}  -  y_{l, n}^{(t)} \big|^{2} +  {v}_{z_{l,n}^{(t)}}  \Big) }.
\end{align}

The BP-MF message passing algorithm (BP-MF-MPA) for active device detection is summarized in Algorithm 1. The estimated hyper-parameter $ \{ \hat{\gamma}_u, \forall u \}$  is used to detect the activity of device $ u $ by comparing it with a threshold $ G_{th} $  (which can be calculated as \cite{Gth2017}), so we have $ \mathsf{\mathcal{I}}_a = \text{find}(\{\hat{\gamma}_u \} < G_{th}) $.

\subsection{Message Passing Algorithm for Non-Coherent Multi-Device Detection}
It is worth mentioning that Fig. \ref{FG5-2} is obtained based on Fig. \ref{FG5-1}, where the variable nodes associated with inactive devices are removed, and the differential modulation constrains are added. BP and MF message passing is used to realize the estimation of $  \{ \psi_{k} , \forall k\} $. The message computations of the function nodes {\small  $  \{ f_{y_{_{l,n}}}^{(t)}( z_{l,n}^{(t)},\lambda_n),\forall l,n\}$} and {\small $ \{ f_{z_{_{l,n}}}^{(t)}, \forall l,n \} $} are similar to those in Fig. \ref{FG5-1}, which are omitted in this section. The function nodes {\small $ \{ f_{\delta_{k,n}}^{(t)} (\bar{x}_{k,n}^{(t)},\bar{x}_{k,n}^{(t-1)},\psi_k^{(t)}), \forall k, n \}  $} are handled by the BP rule, and some messages are approximated to be Gaussian based on moment matching to make the message passing tractable. As the message computation for the $(t-1)$-th measurement is similar to that for the $t$-th, we just detail the message computation for $ t $-th measurement.

\subsubsection{Forward message computations}
Similar to (\ref{eq:fy2zS5}) - (\ref{eq:x2fxM}), we have 
\begin{align}\label{eq:x2fd}
I_{\bar{x}_{k,n}^{(t)} \rightarrow f_{\delta_{k,n}}^{(t)}}\big(\bar{x}_{k,n}^{(t)}\big)  
\propto \mathcal{CN} \left(\bar{x}_{k,n}^{(t)} ; \rarrow{m}_{\bar{x}_{k,n}^{(t)}}, \rarrow{v}_{\bar{x}_{u,n}^{(t)}}\right),
\end{align}
where
\begin{eqnarray}   
\label{eq:Vx2fd}  \rarrow{v}_{\bar{x}_{k,n}^{(t)}}  &\approx& \left(\sum_{l=1}^L \frac{\left| \bar{P}_{_{l, k}} \right|^{2}}{\hat{\lambda}_{n}^{-1}+\larrow{v}_{z_{l, n}^{(t)}}}\right)^{-1},
\end{eqnarray} 
and
\begin{eqnarray}
\label{eq:Mx2fd}
\rarrow{m}_{\bar{x}_{k,n}^{(t)}} &\approx& \rarrow{v}_{\bar{x}_{k,n}^{(t)}} \sum_{l=1}^L \frac{\bar{P}_{_{l, k}}^{H}\left(y_{l, n}^{(t)}-\larrow{m}_{z_{l, n}^{(t)}}\right)}{\hat{\lambda}_{n}^{-1}+\larrow{v}_{z_{l, n}^{(t)}}} + {m}_{\bar{x}_{k,n}}^{(t)}.
\end{eqnarray}
With  {\small $ I_{\bar{x}_{k,n}^{(t)} \rightarrow f_{\delta_{k,n}}^{(t)}}\big(\bar{x}_{k,n}^{(t)}\big) $} and {\small $ I_{\bar{x}_{k,n}^{(t-1)} \rightarrow f_{\delta_{k,n}}^{(t)}}\big(\bar{x}_{k,n}^{(t-1)}\big) $}, the outgoing forward message at $ f_{\delta_{k, n}}^{(t)} $ can be computed as 
\begin{align}
\notag  I_{f_{\delta_{k, n}}^{(t)} \rightarrow \psi_{k}^{(t)}}\big(\psi_{k}^{(t)}\big) & =\int f_{\delta_{k, n}}^{(t)}\big(\bar{x}_{k,n}^{(t)}, \psi_{k}^{(t)}\big) I_{\bar{x}_{k,n}^{(t)} \rightarrow f_{\delta_{k, n}}^{(t)}}\big(\bar{x}_{k,n}^{(t)}\big) \\
& \qquad \quad \cdot I_{\bar{x}_{k,n}^{(t-1)} \rightarrow f_{\delta_{k}, n}^{(t)}}\big(\bar{x}_{k,n}^{(t-1)}\big) d \bar{x}_{k,n}^{(t)} d \bar{x}_{k,n}^{(t-1)} \\
\notag &= \mathcal{CN} \left(\rarrow{m}_{\bar{x}_{k,n}^{(t)}} ; \psi_{k}^{(t)} \rarrow{m}_{\bar{x}_{k,n}^{(t-1)}}, \rarrow{v}_{\bar{x}_{k,n}^{(t)}}+\big|\psi_{k}^{(t)}\big|^{2} \rarrow{v}_{\bar{x}_{k,n}^{(t-1)}}\right).
\end{align}
With {\small $ \{I_{f_{\delta_{k, n}}^{(t)} \rightarrow \psi_{k}^{(t)}}\big(\psi_{k}^{(t)}\big) , \forall n \} $} and the a priori 
{\small $ f_{\psi_{k}}^{(t)}\big(\psi_{k}^{(t)}\big) $}, the belief of $ \psi_{k}^{(t)} $  is given as 
\begin{align}
b\big(\psi_{k}^{(t)}\big)  = f_{\psi_{k}}^{(t)}\big(\psi_{k}^{(t)}\big) \prod_{n = 1}^{N} I_{f_{\delta_{k, n}}^{(t)} \rightarrow \psi_{k}^{(t)}}\big(\psi_{k}^{(t)}\big) 
\triangleq \sum_{q \in \mathcal{X}_d} \beta_{k, t}^{q} \delta\big( \psi_{k}^{(t)}-q \big),
\end{align}
where
\begin{align}\label{eq:BetaS5}
\beta _{k, t}^{q} = \frac{\prod\limits_{n = 1}^{N} \mathcal{CN} \left(\rarrow{m}_{\bar{x}_{k,n}^{(t)}} ; q \rarrow{m}_{\bar{x}_{k,n}^{(t-1)}}, \rarrow{v}_{\bar{x}_{k,n}^{(t)}}+\big| q \big|^{2} \rarrow{v}_{\bar{x}_{k,n}^{(t-1)}}\right)  }{ \sum\limits_{q'\in \mathsf{\mathcal{X}_d} } \prod\limits_{n = 1}^{N}  \mathcal{CN} \left(\rarrow{m}_{\bar{x}_{k,n}^{(t)}} ; q' \rarrow{m}_{\bar{x}_{k,n}^{(t-1)}}, \rarrow{v}_{\bar{x}_{k,n}^{(t)}}+\big| q' \big|^{2} \rarrow{v}_{\bar{x}_{k,n}^{(t-1)}}\right) }  .
\end{align}

\subsubsection{Backward message computations}
With the messages {\small $ \{I_{f_{\delta_{k, n'}}^{(t)} \rightarrow \psi_{k}^{(t)}}\big(\psi_{k}^{(t)}\big) , \forall n' \neq n \} $ } and the  a priori   {\small $ f_{\psi_{k}}^{(t)}\big(\psi_{k}^{(t)}\big) $}, the outgoing backward message at $ \psi_{k}^{(t)} $  can be expressed as 
\begin{align}
\notag I_{\psi_{k}^{(t)} \rightarrow  f_{\delta_{k, n}}^{(t)} }\big(\psi_{k}^{(t)}\big)  & = f_{\psi_{k}}^{(t)}\big(\psi_{k}^{(t)}\big) \prod_{n' \neq n} I_{f_{\delta_{k, n'}}^{(t)} \rightarrow \psi_{k}^{(t)}}\big(\psi_{k}^{(t)}\big) \\
& \triangleq \sum_{q \in \mathcal{X}_d} \alpha_{k, t}^{q} \delta\big( \psi_{k}^{(t)}-q \big),
\end{align}
where
\begin{align}\label{eq:AlphaS5}  
\alpha_{k, t}^{q} = \frac{\prod\limits_{ n' \neq n }  \mathcal{CN} \left(\rarrow{m}_{\bar{x}_{k,n'}^{(t)}} ; q \rarrow{m}_{\bar{x}_{k,n'}^{(t-1)}}, \rarrow{v}_{\bar{x}_{k,n'}^{(t)}}+\big| q \big|^{2} \rarrow{v}_{\bar{x}_{k,n'}^{(t-1)}}\right)  }{ \sum\limits_{q'\in \mathsf{\mathcal{X}}_d} \prod\limits_{n' \neq n}   \mathcal{CN} \left(\rarrow{m}_{\bar{x}_{k,n'}^{(t)}} ; q' \rarrow{m}_{\bar{x}_{k,n'}^{(t-1)}}, \rarrow{v}_{\bar{x}_{k,n'}^{(t)}}+\big| q' \big|^{2} \rarrow{v}_{\bar{x}_{k,n'}^{(t-1)}}\right) } . 
\end{align}
With {\small $ I_{\psi_{k}^{(t)} \rightarrow  f_{\delta_{k, n}}^{(t)} }\big(\psi_{k}^{(t)}\big) $} and {\small $ I_{\bar{x}_{k,n}^{(t-1)} \rightarrow f_{\delta_{k,n}}^{(t)}}\big(\bar{x}_{k,n}^{(t-1)}\big) $}, the outgoing backward message from $ f_{\delta_{k,n}}^{(t)} $ to $ \bar{x}_{k,n}^{(t)} $ can be expressed as
\begin{align}\label{eq:fd2x}
I_{f_{\delta_{k,n}}^{(t)} \rightarrow \bar{x}_{k,n}^{(t)} }\big(\bar{x}_{k,n}^{(t)}\big) 
\notag &= \!\! \int \!\! f_{\delta_{k, n}}^{(t)}\big(\bar{x}_{k,n}^{(t)}, \bar{x}_{k,n}^{(t-1)}, \psi_{k}^{(t)}\big) I_{\bar{x}_{k,n}^{(t-1)} \rightarrow f_{\delta_{k, n}}^{(t)}}\big(\bar{x}_{k,n}^{(t-1)}\big)\\
&\qquad \quad \cdot I_{\psi_{k}^{(t)} \rightarrow f_{\delta_{k, n}}^{(t)}}\big(\psi_{k}^{(t)}\big) d \bar{x}_{k,n}^{(t-1)} d \psi_{k}^{(t)} \\
\notag &=\sum_{q \in \mathcal{X}_d } \alpha_{k, t}^{q}|q|^{2} \mathcal{CN}\left(\bar{x}_{k,n}^{(t)} ; q \rarrow{m}_{\bar{x}_{k,n}^{(t-1)}},|q|^{2} \rarrow{v}_{\bar{x}_{k,n}^{(t-1)}}\right),
\end{align}
and with $ I_{\bar{x}_{k,n}^{(t)} \rightarrow f_{\delta_{k,n}}^{(t)}}\big(\bar{x}_{k,n}^{(t)}\big) $ given in (\ref{eq:x2fd}), the belief of $ \bar{x}_{k,n}^{(t)} $  can be updated by

\begin{align}
b\left(\bar{x}_{k,n}^{(t)}\right) 
\notag & = I_{f_{\delta_{k, n}}^{(t)} \rightarrow \bar{x}_{k,n}^{(t)}}\left(\bar{x}_{k,n}^{(t)}\right)  I_{ \bar{x}_{k,n}^{(t)} \rightarrow f_{\delta_{k, n}}^{(t)}  }\left(\bar{x}_{k,n}^{(t)}\right) \\
& \propto \sum_{q \in \mathcal{X}_d } \rho_{k, t}^{q}  \mathcal{CN}\left(\bar{x}_{k,n}^{(t)} ;  {m}_{k,n}^{(t)}, {v}_{k,n}^{(t)}\right),
\end{align}
where
\begin{align}\label{eq:RhoS5} 
\rho_{k, t}^{q} = \frac{\alpha_{k, t}^{q}|q|^{2} \mathcal{CN}\left(\rarrow{m}_{\bar{x}_{k,n}^{(t)}}; q \rarrow{m}_{\bar{x}_{k,n}^{(t-1)}},\rarrow{v}_{\bar{x}_{k,n}^{(t)}} + |q|^{2} \rarrow{v}_{\bar{x}_{k,n}^{(t-1)}}\right)}{ \sum\limits_{q' \in \mathcal{X}_d } \alpha_{k, t}^{q'}|q'|^{2} \mathcal{CN}\left(\rarrow{m}_{\bar{x}_{k,n}^{(t)}}; q' \rarrow{m}_{\bar{x}_{k,n}^{(t-1)}},\rarrow{v}_{\bar{x}_{k,n}^{(t)}} + |q'|^{2} \rarrow{v}_{\bar{x}_{k,n}^{(t-1)}}\right) },
\end{align}
and
\begin{align}
\label{eq:CoV1}  {v}_{k,n}^{(t)} = \left( \frac{1}{\rarrow{v}_{\bar{x}_{k,n}^{(t)}}} +  \frac{1}{|q|^{2} \rarrow{v}_{\bar{x}_{k,n}^{(t-1)}}} \right) ^{-1}, 
\end{align}
\begin{align}
\label{eq:CoM1}  {m}_{k,n}^{(t)} = {v}_{\bar{x}_{k,n}^{(t)}} \left( \frac{\rarrow{m}_{\bar{x}_{k,n}^{(t)}}}{\rarrow{v}_{\bar{x}_{k,n}^{(t)}}} +  \frac{q \rarrow{m}_{\bar{x}_{k,n}^{(t-1)}} }{|q|^{2} \rarrow{v}_{\bar{x}_{k,n}^{(t-1)}}} \right) .
\end{align}

As {\small $ I_{f_{\delta_{k,n}}^{(t)} \rightarrow \bar{x}_{k,n}^{(t)} }\big(\bar{x}_{k,n}^{(t)}\big)  $}  is not Gaussian, the belief  {\small $ b\big(\bar{x}_{k,n}^{(t)}\big) $} is not Gaussian either. With moment matching, {\small $ b\big(\bar{x}_{k,n}^{(t)}\big) $} is approximated to be Gaussian to make the message tractable. Then we have
\begin{align}\label{eq:GblfX}
\notag b^{\scriptscriptstyle G} \left(\bar{x}_{k,n}^{(t)}\right)  & = \text{Proj}_{G}\left\{ b \left(\bar{x}_{k,n}^{(t)}\right) \right\}\\
& \triangleq \mathcal{CN} \left(\bar{x}_{k,n}^{(t)} ; {m}_{\bar{x}_{k,n}^{(t)}}, {v}_{x_{k,n}^{(t)}}\right) ,
\end{align}
where $ \text{Proj}_{\scriptscriptstyle G}\{*\} $  denotes the operation of Gaussian approximation, and the mean and the variance of $  \bar{x}_{k,n}^{(t)} $ are given as  
\begin{align} 
\label{eq:GblfX-M}  {m}_{\bar{x}_{k,n}^{(t)}} = \sum_{q \in \mathcal{X}_d } \rho_{k, t}^{q} {m}_{k,n}^{(t)}, 
\end{align}
\begin{align} 
\label{eq:GblfX-V}  {v}_{\bar{x}_{k,n}^{(t)}} = \sum_{q \in \mathcal{X}_d } \rho_{k, t}^{q} \Big( |{m}_{k,n}^{(t)}|^2 + {v}_{k,n}^{(t)} \Big).
\end{align}
With {\small $ I_{\psi_{k}^{(t)} \rightarrow  f_{\delta_{k, n}}^{(t)} }\big(\psi_{k}^{(t)}\big) $} and {\small $ I_{\bar{x}_{k,n}^{(t)} \rightarrow f_{\delta_{k,n}}^{(t)} }\big(\bar{x}_{k,n}^{(t)}\big) $}, the outgoing backward message from $ f_{\delta_{k,n}}^{(t)} $ to $ \bar{x}_{k,n}^{(t-1)} $can be expressed as
\begin{align}\label{eq:fd2xt-1}
I_{f_{\delta_{k,n}}^{(t)} \rightarrow \bar{x}_{k,n}^{(t-1)} }\big(\bar{x}_{k,n}^{(t-1)}\big) 
\notag &= \!\! \int \!\! f_{\delta_{k, n}}^{(t)}\big(\bar{x}_{k,n}^{(t)}, \bar{x}_{k,n}^{(t-1)}, \psi_{k}^{(t)}\big) I_{\bar{x}_{k,n}^{(t)} \rightarrow f_{\delta_{k, n}}^{(t)}}\big(\bar{x}_{k,n}^{(t)}\big)\\
\notag &\qquad \quad \cdot I_{\psi_{k}^{(t)} \rightarrow f_{\delta_{k, n}}^{(t)}}\big(\psi_{k}^{(t)}\big) d \bar{x}_{k,n}^{(t)} d \psi_{k}^{(t)} \\
&=\sum_{q \in \mathcal{X}_d } \frac{ \alpha_{k, t}^{q} }{ |q|^{2} } \mathcal{CN}\left(\bar{x}_{k,n}^{(t-1)} ; \frac{ \rarrow{m}_{\bar{x}_{k,n}^{(t)}} }{ q } , \frac{ \rarrow{v}_{\bar{x}_{k,n}^{(t)}} }{ |q|^{2} }   \right).
\end{align}
With {\small $ I_{\bar{x}_{k,n}^{(t-1)} \rightarrow f_{\delta_{k,n}}^{(t)}}\big(\bar{x}_{k,n}^{(t-1)}\big) $ }, the belief of $ \bar{x}_{k,n}^{(t-1)} $  can be updated by
\begin{align}
b\left(\bar{x}_{k,n}^{(t-1)}\right) 
\notag & = I_{f_{\delta_{k, n}}^{(t)} \rightarrow \bar{x}_{k,n}^{(t-1)}}\left(\bar{x}_{k,n}^{(t-1)}\right)  I_{ \bar{x}_{k,n}^{(t-1)} \rightarrow f_{\delta_{k, n}}^{(t)}  }\left(\bar{x}_{k,n}^{(t-1)}\right) \\
& \propto \sum_{q \in \mathcal{X}_d } \rho_{k, t-1}^{q}  \mathcal{CN}\left(\bar{x}_{k,n}^{(t-1)} ;  {m}_{k,n}^{(t-1)}, {v}_{k,n}^{(t-1)}\right),
\end{align}
where $ \rho_{k, t-1}^{q}=\rho_{k, t}^{q} $,
\begin{equation}\label{eq:CoV2} 
{v}_{k,n}^{(t-1)} = \left( \frac{|q|^{2}}{\rarrow{v}_{\bar{x}_{k,n}^{(t)}}} +  \frac{1}{ \rarrow{v}_{\bar{x}_{k,n}^{(t-1)}}} \right) ^{-1}, 
\end{equation}
and
\begin{equation}\label{eq:CoM2} 
{m}_{k,n}^{(t-1)}= {v}_{\bar{x}_{k,n}^{(t-1)}} \left( \frac{ q^H  \rarrow{m}_{\bar{x}_{k,n}^{(t)}}}{\rarrow{v}_{\bar{x}_{k,n}^{(t)}}} +  \frac{\rarrow{m}_{\bar{x}_{k,n}^{(t-1)}} }{ \rarrow{v}_{\bar{x}_{k,n}^{(t-1)}}} \right).
\end{equation}
Thus, {\small $  b^{\scriptscriptstyle G}\big(\bar{x}_{k,n}^{(t-1)}\big) $} can be updated in the same way as (\ref{eq:GblfX})-(\ref{eq:GblfX-V}). 

\begin{algorithm} [!t]
	\caption{ \quad BP-MF-MPA for active device detection and non-coherent multi-device data detection} 
	\label{alg:BP-MF}
	{\bf Input:} $\textbf{Y}^{(t)} $, $\textbf{Y}^{(t-1)} $,$ \textbf{P} $, $\left\lbrace p\left( \gamma_u \right) \right\rbrace, $  $ \{ p ( \psi_{k}^{(t)}) \}  $ \\
	\textbf{Initialize:} $ \forall n, \hat{\lambda}_n=10;\  \forall u, \hat{\gamma}_u=1 ;  \  \forall u,n, {m}_{x_{u,n}^{(t)}} =0 ; \\  \hspace*{0.6 in}  \forall l,n, \larrow{m}_{z_{l,n}^{(t)}}=0, \larrow{v}_{z_{l,n}^{(t)}}\!=\!1,\larrow{m}_{z_{l,n}^{(t-1)}}\!=\!0,\! \larrow{v}_{z_{l,n}^{(t-1)}}\!=\!1$\\  
	\textbf{Active Device Detection}: \\	
	1: \textbf{for} $ i=1:N_{A_{itr}} $   \\	
	2: \hspace*{0.2 in} $ \forall u,n $:  update  $ \rarrow{m}_{x_{u, n}^{(t)}} $, $ \rarrow{v}_{x_{u, n}^{(t)}} $ by (\ref{eq:x2fxM}) and (\ref{eq:x2fxV}); \\	
	3: \hspace*{0.2 in} $ \forall u,n $:  update $  {m}_{x_{u, n}^{(t)}} $, $  {v}_{x_{u, n}^{(t)}} $ by  (\ref{eq:mx}) and (\ref{eq:vx});  \\	
	4: \hspace*{0.2 in} $ \forall l,n $:  update $ \larrow{m}_{z_{l,n}^{(t)}} $, $ \larrow{v}_{z_{l,n}^{(t)}} $ by  (\ref{eq:fz2zM})and (\ref{eq:fz2zV}); \\
	5: \hspace*{0.2 in} $ \forall u $:  update  $ \hat{\gamma}_u$ by  (\ref{eq:gammaHat}); \\
	6: \hspace*{0.2 in} update  $ \hat{\epsilon}$ by  (\ref{eq:Ehat});\\
	7: \hspace*{0.2 in} $ \forall l,n$:  update $  {m}_{z_{l,n}^{(t)}} $, $  {v}_{z_{l,n}^{(t)}} $ by (\ref{eq:BlfzM}) and (\ref{eq:BlfzV});  \\
	8: \hspace*{0.2 in}  $ \forall n  $:  update $ \hat{\lambda}_n  $ by(\ref{eq:LmdHat});\\
	9: \textbf{end for}  \\  
	\textbf{Output:} $ \mathsf{\mathcal{I}}_a = \text{find}(\{\hat{\gamma}_u \} < G_{th}) $. \\
	\vspace{0.01cm}\\  
	\textbf{Non-Coherent Multi-Device Data Detection} \\
	10: \textbf{for}  $ i=1:N_{D_{itr}} $ \\
	11: \hspace*{0.15  in} $ \forall k,n $: update $\rarrow{m}_{\bar{x}_{k,n}^{(t)}} $, $ \rarrow{v}_{\bar{x}_{k,n}^{(t)}} $ by (\ref{eq:Mx2fd}) and (\ref{eq:Vx2fd}), and update \\ \hspace*{0.4 in} $\rarrow{m}_{\bar{x}_{k,n}^{(t-1)}} $, $ \rarrow{v}_{\bar{x}_{k,n}^{(t-1)}} $ similarly;\\
	12: \hspace*{0.15 in} $ \forall k,q $: update $ \beta _{k, t}^{q} $ and $ \alpha_{k, t}^{q} $ by (\ref{eq:BetaS5}) and (\ref{eq:AlphaS5}) respectively; \\	
	13: \hspace*{0.15 in} $ \forall k,q $: update $ \rho_{k, t}^{q} $ by (\ref{eq:RhoS5});	\\
	14: \hspace*{0.15 in} $ \forall k,n $: update  ${m}_{k,n}^{(t)}, {v}_{k,n}^{(t)}$ by  (\ref{eq:CoM1}) and (\ref{eq:CoV1}), and update \\ \hspace*{0.4 in}  ${m}_{k,n}^{(t-1)}, {v}_{k,n}^{(t-1)}$    by (\ref{eq:CoM2}) and (\ref{eq:CoV2}) respectively;	\\
	15: \hspace*{0.15 in}  $ \forall k,n $: update $ {m}_{\bar{x}_{k,n}^{(t)}} $, $ {v}_{\bar{x}_{k,n}^{(t)}}  $ by (\ref{eq:GblfX-M}) and (\ref{eq:GblfX-V}), and update \\ \hspace*{0.4 in}  $ {m}_{\bar{x}_{k,n}^{(t-1)}} $, $ {v}_{\bar{x}_{k,n}^{(t-1)}}  $ similarly;	\\
	16: \hspace*{0.15 in} $ \forall l,n $: update $ \larrow{m}_{z_{l,n}^{(t)}} $, $ \larrow{v}_{z_{l,n}^{(t)}} $ by (\ref{eq:fz2zM-2}) and (\ref{eq:fz2zV-2}), and update  \\ \hspace*{0.42 in} $ \larrow{m}_{z_{l,n}^{(t-1)}} $,$ \larrow{v}_{z_{l,n}^{(t-1)}} $ similarly;\\	
	17: \hspace*{0.15 in} $ \forall l,n$:  update $  {m}_{z_{l,n}^{(t)}} $, $  {v}_{z_{l,n}^{(t)}} $ by (\ref{eq:BlfzM}) and (\ref{eq:BlfzV}), and update \\ \hspace*{0.4 in}  $  {m}_{z_{l,n}^{(t-1)}} $, $ {v}_{z_{l,n}^{(t-1)}} $ similarly ;  \\
	18: \hspace*{0.15 in}  $ \forall n  $:  update $ \hat{\lambda}_n  $ by (\ref{eq:LmdHat}); \\
	19: \textbf{end for} \\	
	{\bf Output:}  {$ \forall k: b(\psi_k^{(t)}) = \sum\limits_{q\in \mathsf{\mathcal{X}_d}}{ \beta _{k, t}^{q} \delta ( \psi_k^{(t)} -q )} $, based on which hard decisions can be made}.	
\end{algorithm}

Similar to (\ref{eq:x2fxS5})-(\ref{eq:fz2zM}), the outgoing backward message at $ f_{z_{_{l,n}}}^{(t)} $ can be computed as
\begin{align}\label{eq:fz2zS5-2}
I_{f_{z_{_{l,n}}}^{(t)} \rightarrow z_{l,n}^{(t)} }\big( z_{l,n}^{(t)}\big)
\propto \mathcal{CN}\left( z_{l,n}^{(t)}; \larrow{m}_{z_{l,n}^{(t)}}, \larrow{v}_{z_{l,n}^{(t)}} \right),
\end{align}
where
\begin{align}\label{eq:fz2zM-2}  
\larrow{v}_{z_{l,n}^{(t)}} \approx \sum_{k = 1}^{K} \left|\bar{P}_{_{l, k}}\right|^{2} {v}_{x_{k, n}^{(t)}} , 
\end{align} 
\begin{align}\label{eq:fz2zV-2}  
\larrow{m}_{z_{l,n}^{(t)}} \approx \sum_{k = 1}^{K} \bar{P}_{_{l, k}} {m}_{x_{k, n}^{(t)}}  - \frac{\larrow{v}_{z_{l,n}^{(t)}} \big( y_{l, n}^{(t)}- {^{i-1} \larrow{m}_{z_{l, n}^{(t)}}} \big) }{ \hat{\lambda}_{n}^{-1} + {^{i-1} \larrow{v}_{z_{l, n}^{(t)}} }}.
\end{align} 
Similar to (\ref{eq:Blfz})-(\ref{eq:fy2lmdS5}), we can calculate  $ \{  I_{f_{y_{_{l,n}}}^{(t)} \rightarrow \lambda_{n} }\big( \lambda_{n} \big)   ,\forall l \} $  and  $ \{  I_{f_{y_{_{l,n}}}^{(t-1)} \rightarrow \lambda_{n} }\big( \lambda_{n} \big)   ,\forall l \} $. With the  a priori   $  f_{\lambda_n }\left( \lambda_n \right) $, the belief of $ \lambda_n $ can be updated by
\begin{align}\label{eq:blfLmd-2}
\notag b \big( \lambda_{n} \big)  & = f_{\lambda_n }\left( \lambda_n \right) \prod_{l = 1}^{L} I_{f_{y_{_{l,n}}}^{(t)} \rightarrow \lambda_{n} }\big( \lambda_{n} \big)  I_{f_{y_{_{l,n}}}^{(t-1)} \rightarrow \lambda_{n} }\big( \lambda_{n} \big)   \\
& \propto \lambda_{n}^{2L-1} \exp \Big\{ \!\! - \lambda_{n}   \sum_{l = 1}^{L} \Big( \big|  {m}_{z_{l,n}^{(t)}}  -  y_{l, n}^{(t)} \big|^{2} +  {v}_{z_{l,n}^{(t)}}  \\ 
\notag & \qquad \qquad \qquad \qquad \qquad +  \big|  {m}_{z_{l,n}^{(t-1)}}  -  y_{l, n}^{(t-1)} \big|^{2} +  {v}_{z_{l,n}^{(t -1)}}  \Big)   \Big\}.
\end{align}
Thus, the estimation of noise precision is given by
\begin{align}\label{eq:LmdHat-2}
\hat{\lambda}_n 
\notag & ={\left\langle \lambda_{n}  \right\rangle}_{b_{\lambda_{n}} \big( \lambda_{n} \big)} \\
& =\frac{ 2L}{  \sum\limits_{l = 1}^{L} \Big(  \big|  {m}_{z_{l,n}^{(t)}}  -  y_{l, n}^{(t)} \big|^{2} +  {v}_{z_{l,n}^{(t)}} +  \big|  {m}_{z_{l,n}^{(t-1)}}  -  y_{l, n}^{(t-1)} \big|^{2} +  {v}_{z_{l,n}^{(t -1)}}  \Big) }.
\end{align}

The BP-MF message passing algorithm (BP-MF-MPA) for multi-device data detection is summarized in Algorithm 1. The output {\small $ \{ b(\psi_k^{(t)}) = \sum_{q\in \mathsf{\mathcal{X}}_d}{ \beta _{k, t}^{q} \delta ( \psi_k^{(t)} -q )},  \forall k \} $} are used for making hard decisions on the symbols, i.e., 
\begin{equation}
\hat \psi_{k} = \mathop{\arg\max}\limits_{\psi_k^{(t)} \in \mathsf{\mathcal{X}}_d } b\left( \psi_{k} \right).
\end{equation}

\subsection{Computational Complexity}
  It can be seen from Algorithm 1 that the complexity of the proposed BP-MF-MPA algorithm is dominated by the multiplication operations to update the mean and variance of the relevant messages. The computational complexity is $ 6UNL+2UN $ in Line 2, $ 4UN $ in Line 3, $ 3UNL+3NL $ in Line 4, $ UN+U $ in Line 5, 2 in Line 6, $ 5NL $ in Line 7, $ NL+N $ in Line 8, $ 12NLK+6NK $ in Line 11, $ 200NK-96K $ in Line 12, $ 164K $ in Line 13, $ 19NK $ in Line 14, $ 24NK $ in Line 15, $ 8NLK+6NL $ in Line 16, $ 10NL $ in Line 17, and $ NL+N $ in Line 18. Thus, the computational complexity for non-coherent multi-device detection can be approximated as $ O(UNL)+ O(KNL) $ per iteration.

\begin{table}[!t]
	\centering 
	\caption{ Simulation Parameters }\label{tab51}
	\begin{tabular}{m{4.5cm} c m{1.5cm}}
		\toprule[1pt] 
		\textbf{Parameter} & \textbf{Symbol} & \textbf{Value} \\
		\midrule[0.5pt]
		Number of devices & $ U $ & 100,300,500 \\
		Number of antennas at AP & $ N $ & 50,100 \\
		Number of active devices & $ K $ &  $ \lfloor 0.1U \rfloor $ \\
		Length of spreading sequences & $ L $ & 11,13\\
		\bottomrule[1pt]
	\end{tabular}
\end{table}

\section{Simulation results}

We assume a MIMO-NOMA system with parameters shown in Table \ref{tab51}, and DQPSK modulation is employed.  The number of active devices is about $10\%$ of the total number of devices, which are randomly selected from the users and access the channel randomly.   

\begin{figure}[t]
	\centering 
	\includegraphics[width=3.5in]{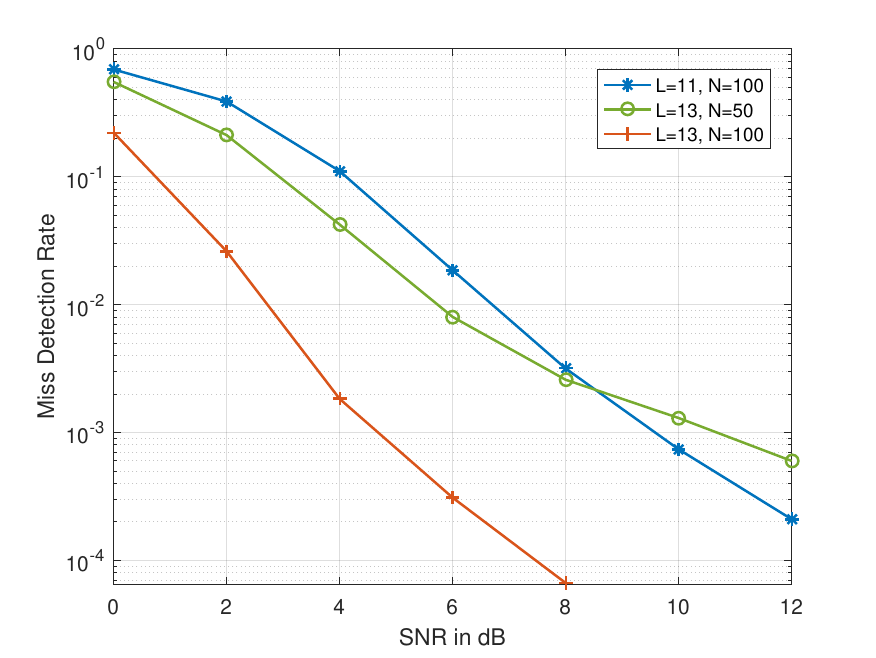}
	\caption{ Missed detection rate performance comparison. }\label{fig:MissR}
\end{figure}

\begin{figure}[t]
	\centering 
	\includegraphics[width=3.5in]{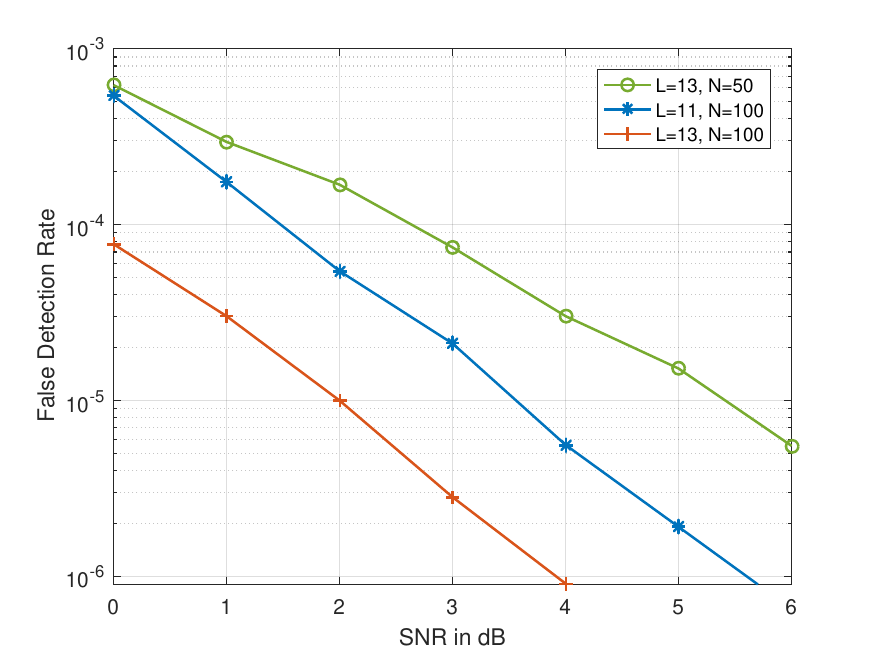}
	\caption{ False detection rate performance comparison. }\label{fig:FR}
\end{figure}

The rate of incorrect detection (miss or false detection) is used to evaluate the performance of active device detection. The miss detection rate and false detection rate of the proposed algorithm are shown in Fig.\ref{fig:MissR} and Fig.\ref{fig:FR} respectively. It can be seen that with a fixed number of antennas at AP $N=100$, the detection performance becomes better with the increase of $ L $, as longer spreading sequences lead to smaller correlations. 
In addition, with a fixed length of spreading sequences $L=13$, we examine the miss detection rate and false detection rate by varying $ N $. As expected, a larger number of antennas lead to a better detection performance as more observations can be obtained by the receiver.
 
\begin{figure}[t]
	\centering 
	\includegraphics[width=3.5in]{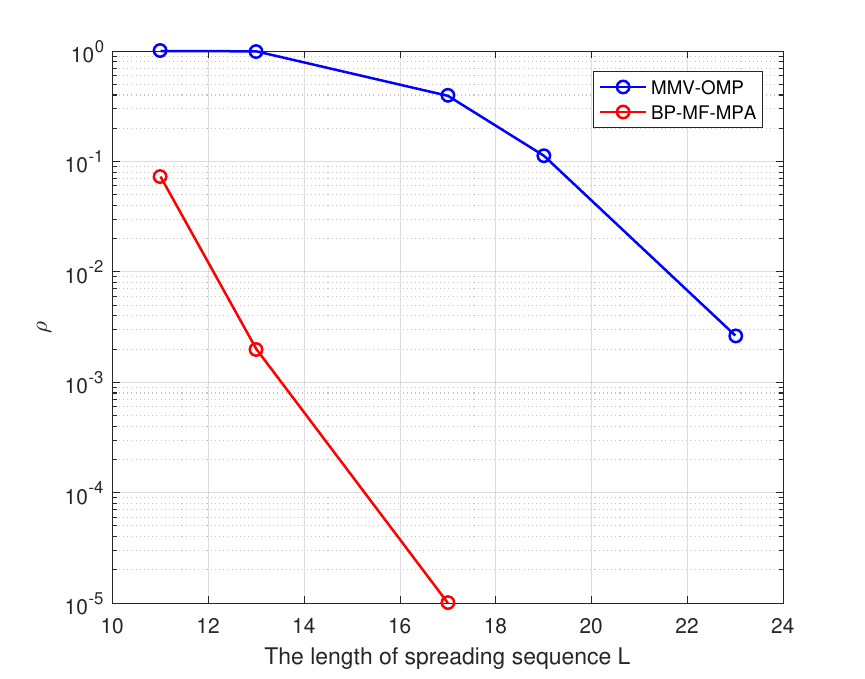}
	\caption{ The support recovery comparison, where $ U=100$ and $ N=50 $.  }\label{fig:SRR}
\end{figure}

In Fig.\ref{fig:SRR}, we evaluate the support recovery rate of the algorithms. As a traditional compressed sensing support recovery algorithm, MMV-OMP is included. We assume the number of devices $ U=100 $, the number of antennas at AP $ N=50 $, and the number of active devices is 10\% of the total devices. The support recovery rate is defined as the percentage of successful trials in the total trials. We defined $ \rho $=1-support recovery rate, and evaluate the $ \rho $ of the algorithms versus the length of the spreading sequences L, i.e., the column size of the spreading matrix P. It turns out that, due to the lack of the mechanism of exploiting common support, the support recovery performance of MMV-OMP is inferior to that of BP-MF-MPA.

\begin{figure}[t]
	\centering 
	\includegraphics[width=3.5in]{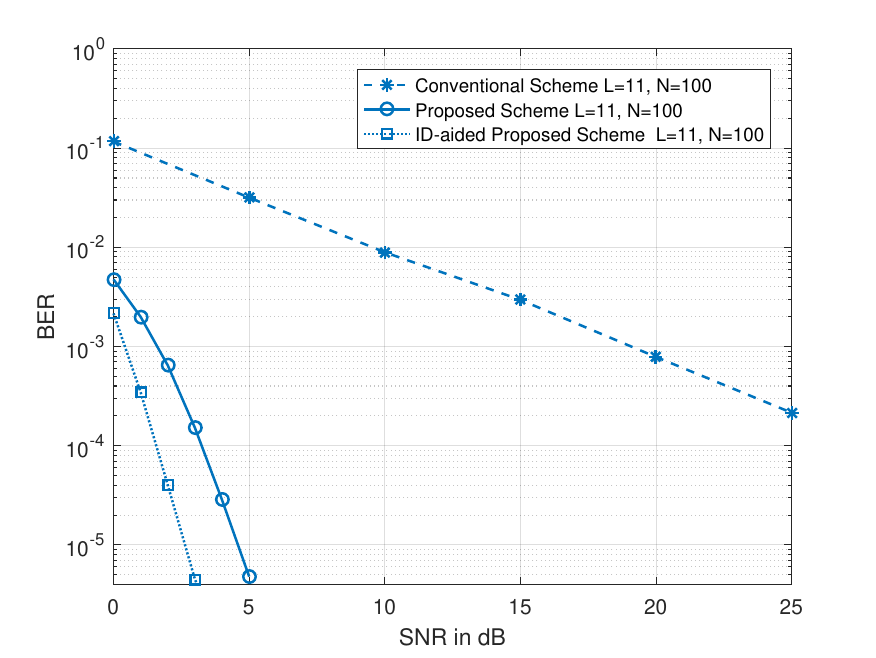}
	\caption{  BER performance comparison of multi-device data detection, where $ L=11 $ and $ N=100 $.  }\label{fig:BERvsMMSE}
\end{figure}

\begin{figure}[t]
	\centering 
	\includegraphics[width=3.5in]{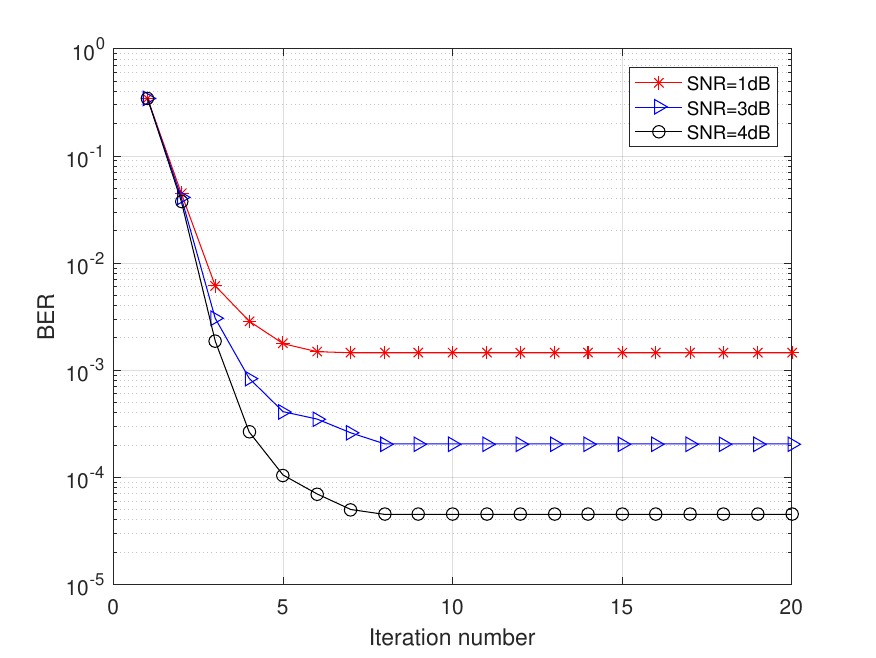}
	\caption{  The convergence of BP-MF-MPA, where $ L=13 $ and $ N=50 $ }\label{fig:cov}
\end{figure} 

\begin{figure}[t]
	\centering 
	\includegraphics[width=3.5in]{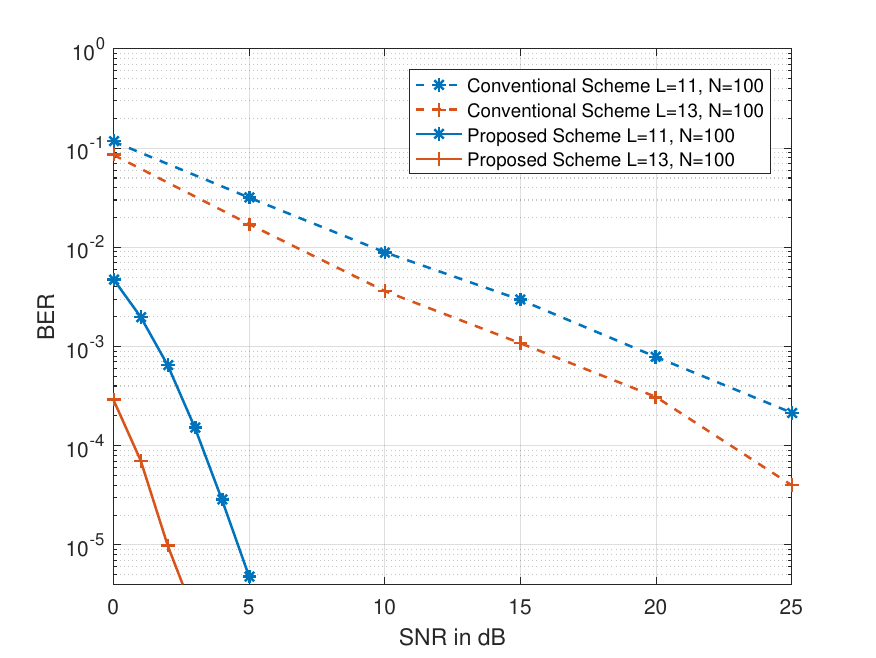}
	\caption{ BER performance comparison of multi-device data detection for different $ L $, where $ N=100 $.  }\label{fig:BERvsL}
\end{figure}

\begin{figure}[t]
	\centering 
	\includegraphics[width=3.5in]{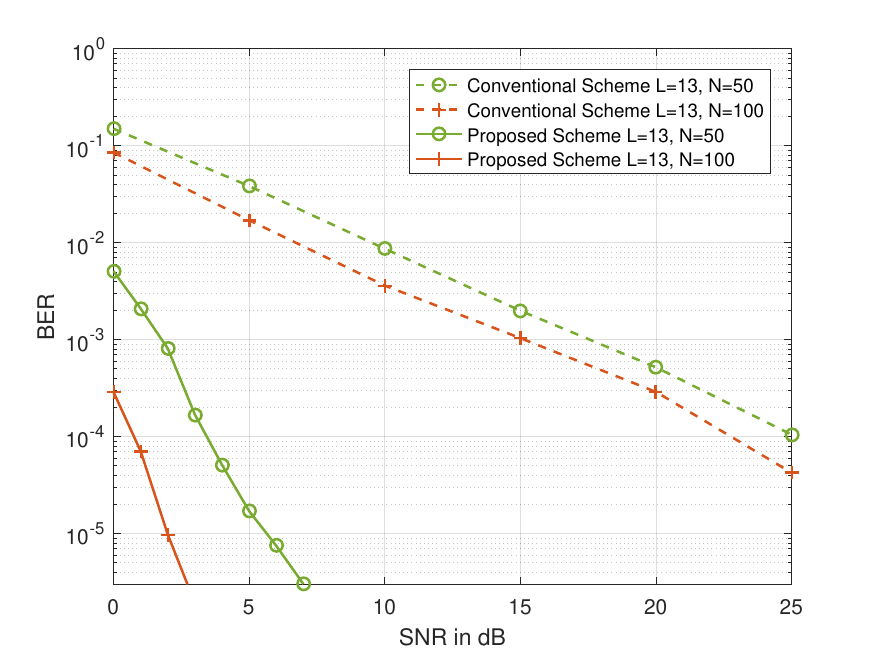}
	\caption{ BER performance comparison of multi-device data detection for different $ N $, where $ L=13 $. }\label{fig:BERvsN}
\end{figure}

\begin{figure}[t]
	\centering 
	\includegraphics[width=3.5in]{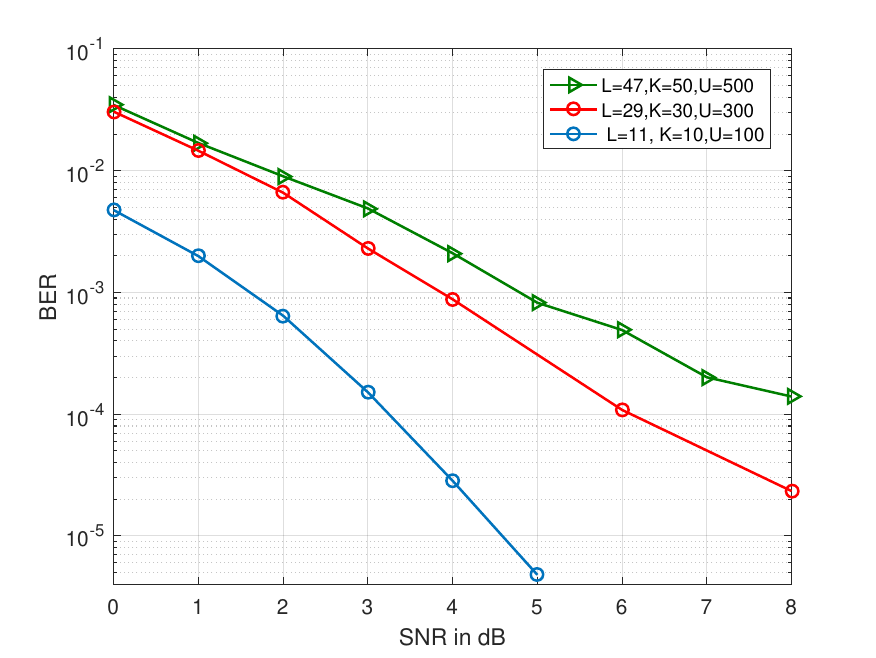}
	\caption{ BER performance comparison with different number of (active) user, where $ N=100 $. }\label{fig:BERvsU}
\end{figure} 

We use the bit error rate (BER) to evaluate the performance of multi-device data detection. To demonstrate the superior performance of the proposed algorithm, which exploits the differential modulation constraint, we compare it with the conventional detection scheme. With the results of active device detection, the inactive devices are removed from the model. Then the estimation of $\bar{\textbf{X}}$ becomes a conventional problem, which can be solved by using the linear minimum mean squared error estimator. The estimates of $\bar{\textbf{X}}$ of two consecutive times can be obtained, which are denoted by      
$\bar{\textbf{X}}^{(t)} $ and $ \bar{\textbf{X}}^{(t-1)}$. Then the demodulated symbol can be expressed as  $ \hat \psi_{k} = 1/N \sum_{n=1}^{N}\bar{x}{_{k,n}}^{(t)} / \bar{x}{_{k,n}}^{(t-1)} $, based on which hard decisions can be made. The proposed scheme is significantly different from the conventional scheme in that the estimation of $\{\psi_{k}, \forall k \}$ is based on $\textbf{Y}^{(t-1)}$ and $ \textbf{Y}^{(t)}$ directly and jointly.

The BER performance of the proposed scheme and the conventional scheme with respect to signal-to-noise ratios (SNRs) is shown in Fig. \ref{fig:BERvsMMSE},
where the length of spreading sequence $ L=11 $ and the number of antennas at AP $ N=100 $. As a performance benchmark, the ID-aided MPA scheme, where the common support is perfectly known, is also included.  It can be seen that the gap between the proposed scheme and the ID-aided MPA scheme is about 2dB at the BER of $ 10^{-5} $.
In addition, compared with the conventional scheme, the proposed scheme can improve the BER performance significantly over the entire SNR range, because the proposed BP-MF-MPA algorithm make full use of the differential constrain information. 

The convergence of the proposed scheme for non-coherent multi-device data detection is shown in Fig.\ref{fig:cov}. It can be seen that the BER of BP-MF-MPA falls rapidly within the first 5 iterations, and it converges with about Nitr = 10. One of the advantage of the message passing algorithms is that they can be implemented in parallel if multiple processing units are available, so that the time required for convergence can be reduced.  

With different  $ L $, the BER performance of the two schemes is shown in Fig. \ref{fig:BERvsL}. It can be observed that, the schemes with $ L=13 $ delivers better BER performance than those with $ L=11 $. This is because longer spreading sequence can reduce the cross-correlation of the spreading sequences, which brings down the multi-user interference. Finally, we examine the BER performance of the schemes by varying the number of antennas $ N $ equipped at AP, and the
results are shown in Fig. \ref{fig:BERvsN}. It can be seen that, with the increase of $ N $, the BER performance is improved as expected.  

The BER performance with different number of (active) users is shown in Fig. \ref{fig:BERvsU}. We assume that the number of antennas at AP $ N=100 $, and the number of users $ U = $ 100, 300 and 500 , and the active users is 10\% of the total users. It turns out that, as more users will lead to stronger multi-user interference, the BER performance has decreased with the increase of the number of (active) users. But the proposed scheme is still able to handle the multi-user interference effectively.

\section{Conclusion}
In this paper, we have investigated the design of grant-free MIMO-NOMA system for machine type communications with the aim to achieve high efficiency and low latency. In particular, differential modulation is employed so that the costly channel estimation in MIMO-NOMA can be bypassed, and the use of long training signals are avoided. The receiver needs to perform active device detection and multi-device detection. We have designed message passing based SBL algorithm to solve the active device detection problem. We have also developed message passing based non-coherent multi-device data detector, where the differential modulation is exploited as a constraint in the detection, leading to significant performance improvement compared to conventional demodulation scheme. Both active device detection and multi-device data detection are performed symbol-by-symbol, which allow highly flexible transmission of the IoT devices. Simulation results verified the effectiveness of the proposed detectors.

\appendices
\section{Derivations of (\ref{eq:x2fxM}) and (\ref{eq:fz2zM})}

\subsection {Derivation of  (\ref{eq:x2fxM})}

\begin{small}
	\begin{equation*}
	\begin{split}
	\rarrow{m}_{x_{u, n}^{(t)}} &= \rarrow{v}_{x_{u, n}^{(t)}} \left(\sum_{l=1}^L \frac{ \overset{\scriptscriptstyle\twoheadrightarrow}{m}_{x_{u, n,l}^{(t)}} } { \overset{\scriptscriptstyle\twoheadrightarrow}{v}_{x_{u, n,l}^{(t)}} }  \right) \\
	&\approx  \rarrow{v}_{x_{u, n}^{(t)}} \sum_{l=1}^L  \frac{P_{_{l, u}}^{H}\left(y_{l, n}^{(t)} - \larrow{m}_{z_{l, n}^{(t)}} + P_{_{l, u}} \overset{\scriptscriptstyle\twoheadleftarrow}{m}_{x_{u, n,l}^{(t)}}  \right)}{\hat{\lambda}_{n}^{-1}+\larrow{v}_{z_{l, n}^{(t)}}} \\
	& \overset{(\ref{eq:x2fzM})}{\approx} \rarrow{v}_{x_{u, n}^{(t)}} \sum_{l=1}^L \frac{P_{_{l, u}}^{H}\left(y_{l, n}^{(t)} - \larrow{m}_{z_{l, n}^{(t)}}  \right)}{\hat{\lambda}_{n}^{-1}+\larrow{v}_{z_{l, n}^{(t)}}} \\ 
	& \quad \qquad + \rarrow{v}_{x_{u, n}^{(t)}} \sum_{l=1}^L  \frac{\left|P_{_{l, u}}\right|^{2}\left( {m}_{x_{u, n}^{(t)}} -  {v}_{x_{u, n}^{(t)}} \frac{\overset{\scriptscriptstyle\twoheadrightarrow}{m}_{x_{u, n,l}^{(t)}}}{ \overset{\scriptscriptstyle\twoheadrightarrow}{v}_{x_{u, n,l}^{(t)}} } \right)}{\hat{\lambda}_{n}^{-1}+\larrow{v}_{z_{l, n}^{(t)}}} \\ 
	& \overset{(\ref{eq:x2fxV})(\ref{eq:x2fzV})}{\approx}  \rarrow{v}_{x_{u, n}^{(t)}} \sum_{l=1}^L  \frac{P_{_{l, u}}^{H}\left(y_{l, n}^{(t)} - \larrow{m}_{z_{l, n}^{(t)}}  \right)}{\hat{\lambda}_{n}^{-1}+\larrow{v}_{z_{l, n}^{(t)}}} +  {m}_{x_{u, n}^{(t)}}\\ 
	& \quad \qquad - \rarrow{v}_{x_{u, n}^{(t)}} \sum_{l=1}^L  \frac{\left|P_{_{l, u}}\right|^{2}  \overset{\scriptscriptstyle\twoheadrightarrow}{m}_{x_{u, n,l}^{(t)}}  }{\hat{\lambda}_{n}^{-1}+\larrow{v}_{z_{l, n}^{(t)}}} \\ 
	& \quad \approx \rarrow{v}_{x_{u, n}^{(t)}} \sum_{l=1}^L \frac{P_{_{l, u}}^{H}\left(y_{l, n}^{(t)}-\larrow{m}_{z_{l, n}^{(t)}}\right)}{\hat{\lambda}_{n}^{-1}+\larrow{v}_{z_{l, n}^{(t)}}} + {m}_{x_{u,n}^{(t)}}.
	\end{split}
	\end{equation*}	
\end{small}

\subsection {Derivation of (\ref{eq:fz2zM})}
\begin{small}
	\begin{equation*}
	\begin{split}
	\larrow{m}_{z_{l,n}^{(t)}} &= \sum_{u = 1}^{U} P_{_{l, u}} \overset{\scriptscriptstyle\twoheadleftarrow}{m}_{x_{u, n,l}^{(t)}}\\ 
	& \overset{(\ref{eq:x2fzM})}{\approx} \sum_{u = 1}^{U} P_{_{l, u}} \left( {m}_{x_{u, n}^{(t)}} -  {v}_{x_{u, n}^{(t)}} \frac{\overset{\scriptscriptstyle\twoheadrightarrow}{m}_{x_{u, n,l}^{(t)}}}{ \overset{\scriptscriptstyle\twoheadrightarrow}{v}_{x_{u, n,l}^{(t)}} }\right)  \\
	& \overset{(\ref{eq:fz2xM}(\ref{eq:fz2xV})}{\approx} \sum_{u = 1}^{U} \left( P_{_{l, u}} {m}_{x_{u, n}^{(t)}} - \frac{  {v}_{x_{u, n}^{(t)}} \left|P_{_{l, u}}\right|^{2}  \big( y_{l, n}^{(t)}- {^{i-1} \larrow{m}_{z_{l, n}^{(t)}}} \big)} {\hat{\lambda}_{n}^{-1} + {^{i-1} \larrow{v}_{z_{l, n}^{(t)}} }} \right) \\
	&\  \overset{(\ref{eq:fz2zV})}{\approx} \sum_{u = 1}^{U} P_{_{l, u}} {m}_{x_{u, n}^{(t)}}  - \frac{\larrow{v}_{z_{l,n}^{(t)}} \big( y_{l, n}^{(t)}- {^{i-1} \larrow{m}_{z_{l, n}^{(t)}}} \big) }{ \hat{\lambda}_{n}^{-1} + {^{i-1} \larrow{v}_{z_{l, n}^{(t)}} }} .
	\end{split}
	\end{equation*}
\end{small}

\bibliographystyle{IEEEtran}% plain unsrt alpha abbrv reference
\bibliography{Bib}

\begin{IEEEbiography}[{\includegraphics[width=1in,height=1.25in,clip,keepaspectratio]{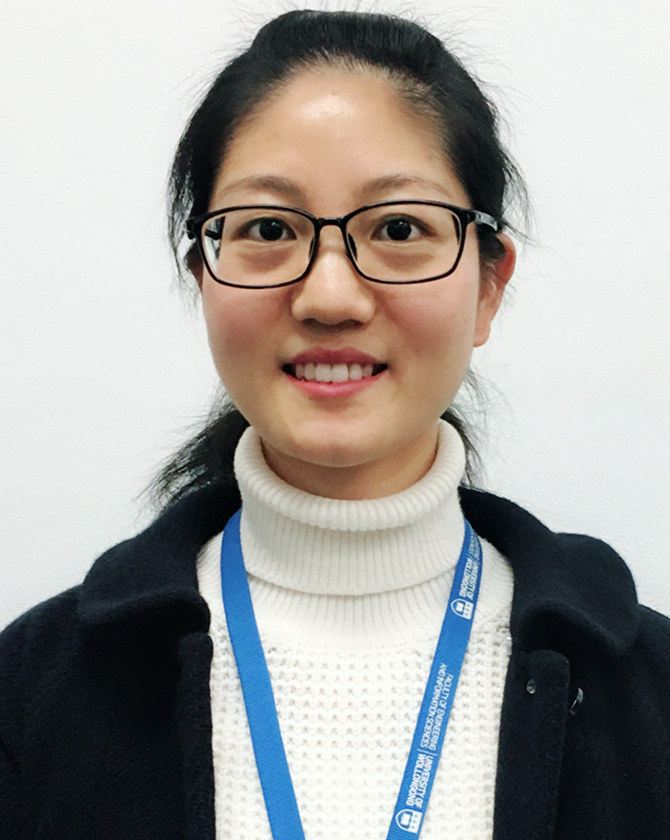}}]{Yuanyuan Zhang}
	received the B.E. degree in electronic information engineering, the M.E. degree and the Ph.D. degree in communication and information systems from Zhengzhou University, Zhengzhou, China, in 2012, 2015 and 2020, respectively.  In 2017-2019, she was a joint Ph.D. student with the School of Electrical, Computer and Telecommunications Engineering, University of Wollongong, Wollongong, NSW, Australia. She is currently a lecturer with the Zhengzhou University of Light Industry.  Her research interests include signal processing in wireless communication networks.
\end{IEEEbiography}

\begin{IEEEbiography}[{\includegraphics[width=1in,height=1.25in,clip,keepaspectratio]{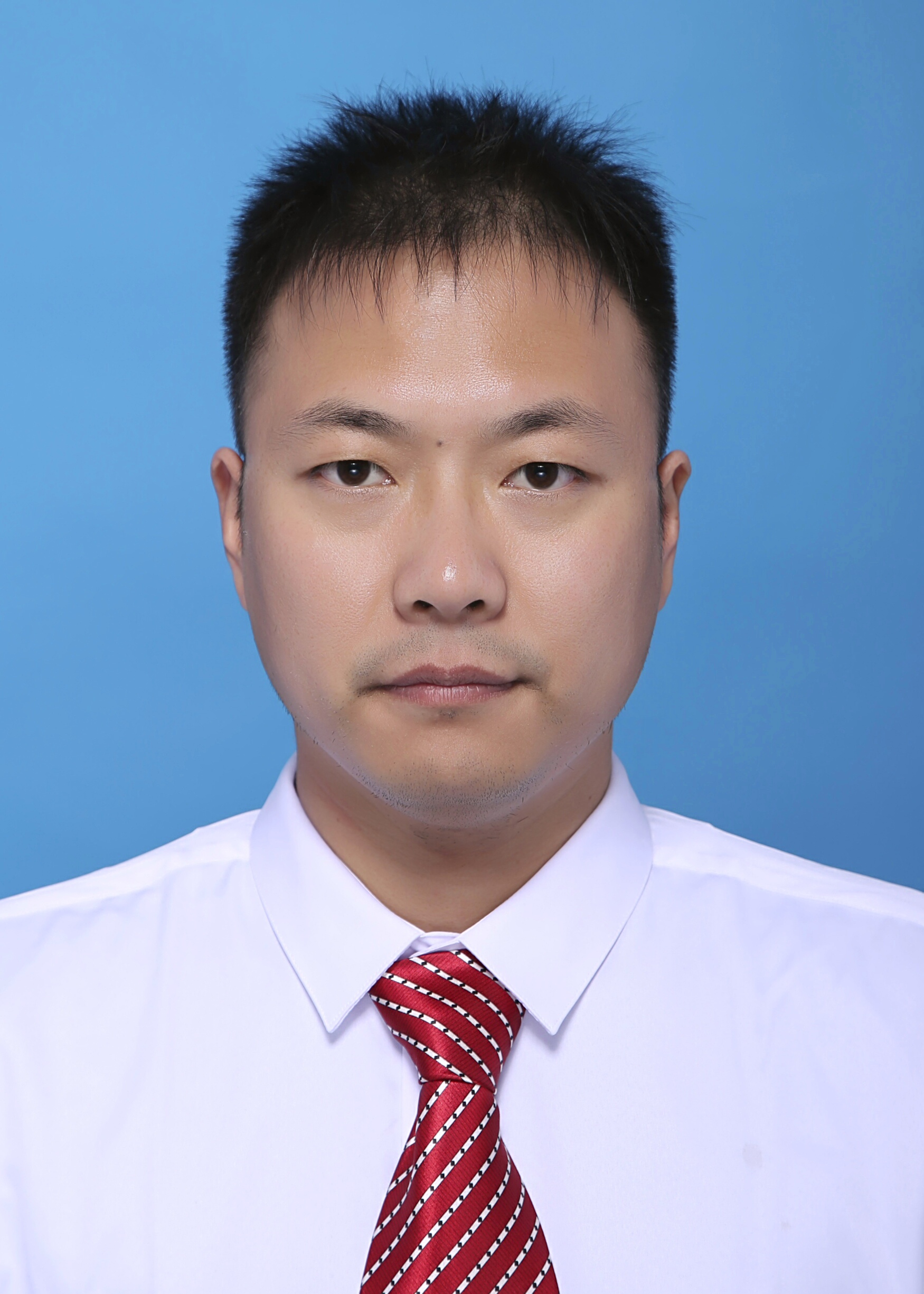}}]{Zhengdao Yuan}
	received the B.E. degree in communication and information system from the Henan University of Science and Technology in 2006, and the M. E. degree in communication engineering from Soochow University in 2009, and the Ph.D. degree in information and communication engineering from the National Digital Switching System Engineering and Technological Research Center in 2018. He is currently an associate professor with the Open University of Henan. He was a visiting scholar with the University of Wollongong in 2019. His research interests are mainly in massive MIMO, sparse channel estimation, message passing algorithms, and iterative receiver design.
\end{IEEEbiography}

\begin{IEEEbiography}[{\includegraphics[width=1in,height=1.25in,clip,keepaspectratio]{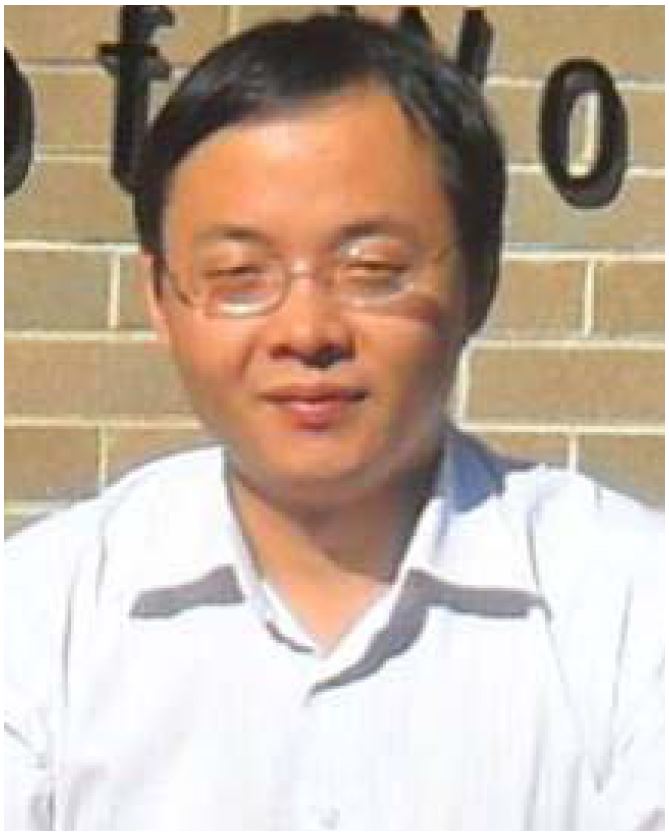}}]{Qinghua Guo}
	(S’07–M’08–SM’18) received the B.E. degree in electronic engineering and the M.E. degree in signal and information processing from Xidian University, in 2001 and 2004, respectively, and the Ph.D. degree in electronic engineering from the City University of Hong Kong in 2008. He is currently an Associate Professor with the School of Electrical, Computer and Telecommunications Engineering, University of Wollongong, Wollongong, NSW, Australia, and an Adjunct Associate Professor with the School of Engineering, The University of Western Australia, Perth, WA, Australia. His research interests include signal processing, telecommunications, radar and optical sensing. He was a recipient of the Australian Research Council’s inaugural Discovery Early Career Researcher Award in 2012.
\end{IEEEbiography}

\begin{IEEEbiography}[{\includegraphics[width=1in,height=1.25in,clip,keepaspectratio]{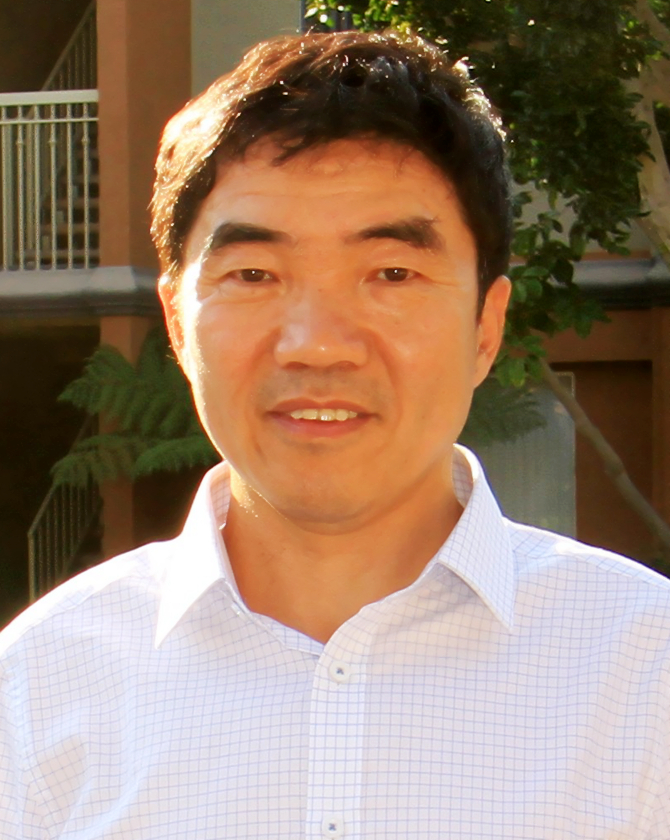}}]{Zhongyong Wang}
	received the B.S. and M.S. degrees in automatic control from the Harbin Shipbuilding Engineering Institute, Harbin, China, in 1986 and 1988, respectively, and the Ph.D. degree in automatic control theory and application from Xi’an Jiaotong University, Xi’an, China, in 1998. Since 1988, he has been a Lecturer with the Department of Electronics, Zhengzhou University, Zhengzhou, China. From 1999 to 2002, he was an Associate Professor. In 2002, he was promoted to Professor with the Department of Communication Engineering. His research interests include numerous aspects of embedded systems, signal processing, and communication theory.
\end{IEEEbiography}

\begin{IEEEbiography}[{\includegraphics[width=1in,height=1.25in,clip,keepaspectratio]{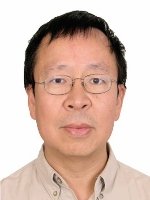}}]{Jiangtao Xi}
	received the B.E. degree in electrical engineering from the Beijing Institute of Technology, Beijing, China, in 1982, the M.E. degree in electrical engineering from Tsinghua University, Beijing, in 1985, and the Ph.D. degree in electrical engineering from the University of Wollongong, Wollongong, NSW, Australia, in 1996. He was a Postdoctoral Fellow with the Communications Research Laboratory, McMaster University, Hamilton, ON, Canada, from 1995 to 1996, and a Member of Technical Staff with Bell Laboratories, Lucent Technologies Inc., Murray Hill, NJ, USA, from 1996 to 1998. From 2000 to 2002, he was the Chief Technical Officer with TCL IT Group Co., China. In 2003, he rejoined the University of Wollongong as a Senior Lecturer, where he is currently a Professor and also the Associate Dean (Global Engagement) of the Faculty of Engineering and Information Sciences. His research interests include signal processing and its applications in various areas, such as optoelectronics, instrumentation and measurement, and telecommunications.
\end{IEEEbiography}

\begin{IEEEbiography}[{\includegraphics[width=1in,height=1.25in,clip,keepaspectratio]{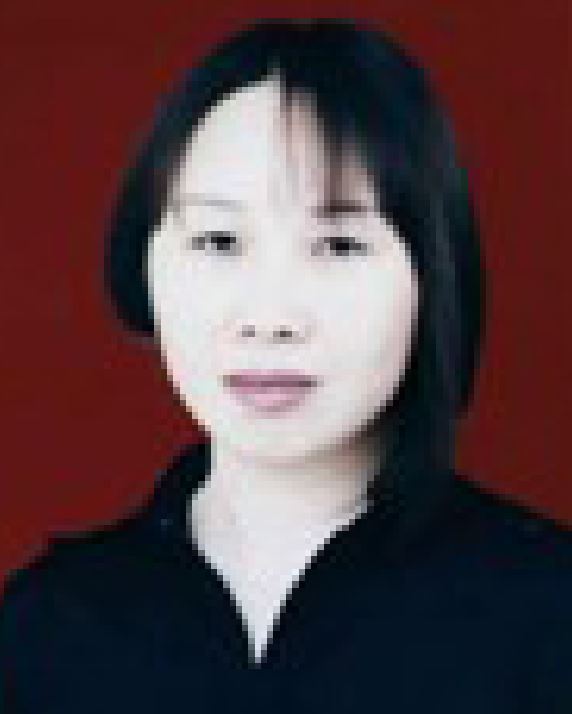}}]{Yanguang Yu}
	received the B.E. degree from the Huazhong University of Science and Technology, Wuhan, China, in 1986, and the Ph.D. degree from the Harbin Institute of Technology, Harbin, China, in 2000. She joined the University of Wollongong, Wollongong, NSW, Australia, in 2007, where she is currently an Associate Professor with the School of Electrical, Computer and Telecommunications Engineering. She was with the College of Information Engineering, Zhengzhou University, Zhengzhou, China, on various appointments, including a Lecturer from 1986 to 1999, an Associate Professor from 2000 to 2004, and a Professor from 2005 to 2007. From 2001 to 2002, she was a Postdoctoral Fellow with Opto-Electronics Information Science and Technology Laboratory, Tianjin University, Tianjin, China. She also had a number of visiting appointments, including a Visiting Fellow with the Optoelectronics Group, Department of Electronics, The University of Pavia, Pavia, Italy, from 2002 to 2003, a Principal Visiting Fellow with the University of Wollongong, from 2004 to 2005, and a Visiting Associate Professor and Professor with the Engineering School ENSEEIHT, Toulouse, France, in 2004 and 2006, respectively. Her research interests include semiconductor lasers with optical feedback and their applications in sensing and instrumentations, and secure chaotic communications. She is also interested in signal processing and its applications to 3-D profile measurement and telecommunication systems.
\end{IEEEbiography}

\begin{IEEEbiography}[{\includegraphics[width=1in,height=1.25in,clip,keepaspectratio]{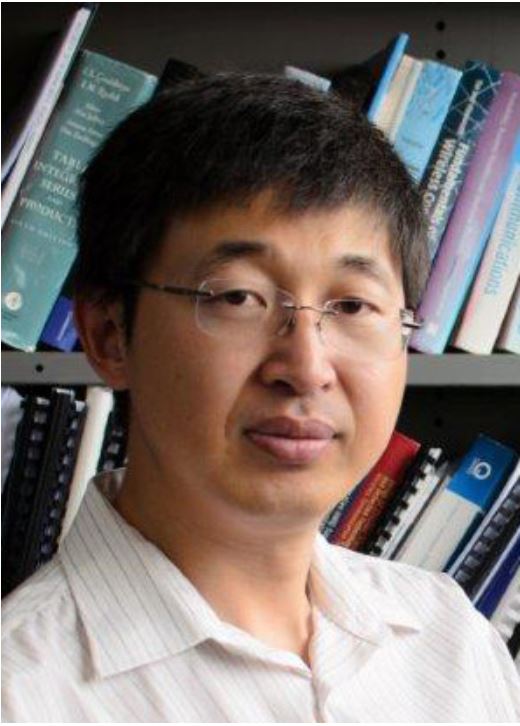}}]{Yonghui Li}
	 (M’04-SM’09-F19) received his PhD degree in November 2002 from Beijing University of Aeronautics and Astronautics. Since 2003, he has been with the Centre of Excellence in Telecommunications, the University of Sydney, Australia. He is now a Professor and Director of Wireless Engineering Laboratory in School of Electrical and Information Engineering, University of Sydney. He is the recipient of the Australian Queen Elizabeth II Fellowship in 2008 and the Australian Future Fellowship in 2012. He is a Fellow of IEEE.	His current research interests are in the area of wireless communications, with a particular focus on MIMO, millimeter wave communications, machine to machine communications, coding techniques and cooperative communications. He holds a number of patents granted and pending in these fields. He was an editor for IEEE transactions on communications and IEEE transactions on vehicular technology. He also served as the guest editor for several IEEE journals, such as IEEE JSAC, IEEE Communications Magazine, IEEE IoT journal, IEEE Access. He received the best paper awards from IEEE International Conference on Communications (ICC) 2014, IEEE PIRMC 2017 and IEEE Wireless Days Conferences (WD) 2014.
\end{IEEEbiography}

\end{document}